\documentclass[sigconf]{acmart}
\usepackage{subfigure} 
\usepackage[ruled]{algorithm2e}
\usepackage{algorithmic}
\usepackage{threeparttable}
\usepackage{wasysym}
\usepackage{enumitem}
\usepackage{multirow}
\usepackage{balance}
\usepackage[table]{xcolor}
\AtBeginDocument{%
  }

\copyrightyear{2026}
\acmYear{2026}
\setcopyright{cc}
\setcctype{by}
\acmConference[SIGIR '26]{Proceedings of the 49th International ACM SIGIR Conference on Research and Development in Information Retrieval}{July 20--24, 2026}{Melbourne, VIC, Australia}
\acmBooktitle{Proceedings of the 49th International ACM SIGIR Conference on Research and Development in Information Retrieval (SIGIR '26), July 20--24, 2026, Melbourne, VIC, Australia}
\acmDOI{10.1145/3805712.3809600}
\acmISBN{979-8-4007-2599-9/2026/07}

\begin{document}

\title{ProMax: Exploring the Potential of LLM-derived Profiles with Distribution Shaping for Recommender Systems}

\author{Yi Zhang}
\authornote{The work was done while the author was visiting the University of Queensland.}
\affiliation{%
  \institution{Anhui University}
  \city{Hefei}
  \country{China}}
\email{zhangyi.ahu@gmail.com}

\author{Yiwen Zhang}
\authornote{Yiwen Zhang and Hongzhi Yin are co-corresponding authors.}
\affiliation{%
  \institution{Anhui University}
  \city{Hefei}
  \country{China}
}
\email{zhangyiwen@ahu.edu.cn}

\author{Kai Zheng}
\affiliation{%
  \institution{University of Electronic Science and Technology of China}
  \city{Chengdu}
  \country{China}
}
\email{zhengkai@uestc.edu.cn}

\author{Tong Chen}
\affiliation{%
  \institution{The University of Queensland}
  \city{Brisbane}
  \country{Australia}
}
\email{tong.chen@uq.edu.au}

\author{Hongzhi Yin}
\authornotemark[2]
\affiliation{%
  \institution{The University of Queensland}
  \city{Brisbane}
  \country{Australia}
}
\email{h.yin1@uq.edu.au}

\renewcommand{\shortauthors}{Yi Zhang, Yiwen Zhang, Kai Zheng, Tong Chen, \& Hongzhi Yin}

\begin{abstract}
The remarkable text understanding and generation capabilities of large language models (LLMs) have revitalized the field of general recommendation based on implicit user feedback. Rather than deploying LLMs directly as recommendation models, a more flexible paradigm leverages their ability to interpret users’ historical interactions and semantic contexts to extract structured profiles that characterize user preferences. These profiles can be further transformed into actionable high-dimensional representations, serving as powerful signals to augment and strengthen recommendation models. However, the mechanism by which such profiles enhance recommendation performance within the feature space remains insufficiently understood. Moreover, existing studies predominantly rely on nonlinear alignment and fusion strategies to incorporate these profiles, which often lead to semantic loss and fail to fully exploit their potential. To address these limitations, we revisit profiles from a retrieval perspective and propose a simple yet effective recommendation framework built upon distribution shaping (\textsf{ProMax}) in this paper. We begin by employing dense retrieval to uncover the collaborative relationships between user and item profiles within the feature space. Based on this insight, we introduce a dual distribution-reshaping process, in which the profile distribution acts as a guiding signal to steer the recommendation model toward learning user preferences for unseen items beyond the scope of observed interactions.
We apply \textsf{ProMax} to four classic recommendation methods on three public datasets. The results indicate that \textsf{ProMax} substantially improves base model performance and outperforms existing LLM-based recommendation approaches.
\end{abstract}



\maketitle

\section{Introduction}
General recommender systems \cite{ricci2011introduction}, widely applied in domains such as web search and E-commerce, typically rely on collaborative filtering (CF) techniques \cite{he2017neural} that exploit ID-level co-occurrence signals derived from historical interactions \cite{yuan2023go}. The prevailing paradigm in this line of research is to learn expressive user and item embedding representations within a unified latent space, thereby providing direct guidance for optimal user–item matching \cite{yin2015joint}. Inference under this paradigm is highly efficient, often requiring only a single forward computation, such as an inner product or a shallow non-linear projection \cite{rendle2020neural}. Nevertheless, realizing this objective is far from trivial, particularly in scenarios where the available observations are insufficient to support estimation \cite{yu2022graph}.

Pre-trained large language models (LLMs) \cite{chang2024survey, zhao2023survey}, empowered by vast textual corpora and strong generative abilities, have recently gained traction in recommender systems as a prospective solution to the sparsity issue \cite{wu2024survey, wang2026mllmrec}. One research line treats LLMs as deep recommenders by embedding historical interactions into prompts for iterative fine-tuning \cite{liao2024llara, gao2025sprec}, yet such approach often amplifies the trade-off between timeliness and training complexity. Alternatively, LLMs act as semantic enhancers \cite{xi2024towards, wei2024llmrec}, distilling user profiles into vectorized representations \cite{wang2025lettingo} that are seamlessly integrated into recommendation models. Representative work such as RLMRec \cite{ren2024representation} builds user profiles from item descriptions and sampled user reviews, where the resulting high-dimensional embeddings are used to enhance model consistency. Subsequent studies extend this paradigm with richer profile structures \cite{wang2025intent}, denoising \cite{wang2025unleashing}, and techniques such as embedding whitening \cite{zhang2024id} and compression \cite{hu2025alphafuse}. Some recent efforts move further by replacing randomly initialized ID embeddings with semantic representations \cite{sheng2025language}. 

These emerging studies collectively highlight the effectiveness of LLM-derived profiles in recommender systems. However, their underlying role remains largely unexplored, leaving open the question of how such opaque profiles truly enhance recommendation models. In general, ID embeddings in recommender models are derived from modeling users’ historical interactions, where the recommendation objective enforces the target user to be close to positive samples \cite{rendle2009bpr} (\textit{i.e.}, items the user has interacted with) in the feature space \cite{yu2022graph}. In contrast, LLM-derived profiles are grounded in textual information, capturing user preferences in a more abstract and generalized manner rather than pointing to any specific item \cite{ren2024representation, xu2026multi}. More importantly, they originate from a pre-trained linguistic space \cite{zhang2022effect} rather than the semantic space shaped within recommendation scenarios. Therefore, the link between ID embeddings and profile representations is intuitively weak, motivating many studies to resort to non-linear networks for alignment \cite{ren2024representation, wang2025intent}, or even directly replace ID embeddings with item's titles or profiles \cite{sheng2025language}. Such additional designs incur extra human effort and training overhead, while inevitably introducing semantic inconsistency and loss for downstream recommendation tasks \cite{sobal2024mathbb, hu2025alphafuse}.

On the other hand, employing profiles merely as auxiliary feature augmentations within recommendation models may fail to fully exploit their potential. Concretely, the practice of using profiles only as auxiliary features reduces high-dimensional semantics to peripheral signals (\textit{e.g.}, forcing dimensional alignment between profile and ID embeddings via MLPs \cite{ren2024representation}), thereby undermining the rich expressiveness of user and item profiles, particularly their cross-item generality and abstraction. Consequently, the recommendation model remains largely bound to sparse interaction patterns. Furthermore, user profiles are inherently high-dimensional and abstract \cite{ren2024representation} (\textit{e.g.}, reflecting a preference for science fiction movies), but often lack the granularity required to differentiate among specific items (\textit{e.g.}, which specific science fiction movie). As a result, directly replacing ID embeddings with profile representations risks producing recommendations that are generic rather than precise. More critically, redundancy and noise in profile information may contaminate training, depriving the recommender model of the opportunity to capture clean interaction-driven signals.

Therefore, we take a different perspective, seeking to answer two central questions: \textit{how can we interpret the beneficial impact of profiles on recommendation, and how can their potential be fully exploited?} To address these two questions, we propose \textbf{\textsf{ProMax}}, a framework built upon LLM-driven user and item profiles that can be seamlessly integrated into mainstream recommendation models. It requires only two additional training objectives while leaving the original model architecture intact. Specifically, we first revisit these LLM-generated profiles from a retrieval perspective to demonstrate that they embed remarkable collaborative signals \cite{wang2019neural}, which fundamentally explain their ability to enhance base recommendation models. Therefore, LLM-driven profiles are qualified to serve as guidance for optimizing recommendation models, without the need to be re-integrated into the models for secondary training.

To maximize the potential of these user and item profiles, we apply them to both supervised distribution reshaping and self-supervised distribution reshaping processes. The former treats the distribution of profile representations as an indicator, employing uncertainty-based weighting to encourage the recommendation model to move beyond the observed items, thereby providing more possibilities for potential unobserved items. In contrast, the latter addresses the risk of the former’s over-dependence on original profiles by reinvigorating users’ historical interactions through profile-based retrieval and LLM-enhanced re-ranking. It further strengthens the expressive capacity of the recommendation model by enforcing bidirectional consistency maximization. The major contributions of this paper are summarized as follows:

\begin{itemize}[leftmargin=*]
\item[$\bullet$] We revisit LLM-driven profiles from the perspective of retrieval and propose \textsf{ProMax}, a simple yet effective model-agnostic framework that exploits profile-guided retrieval to fully uncover users’ latent preferences for unobserved items.
\item[$\bullet$] We propose both supervised distribution reshaping and self-supervised distribution reshaping paradigms to comprehensively exploit the potential of LLM-driven profiles in recommendation.
\item[$\bullet$] We apply \textsf{ProMax} to four classic recommendation methods on three public datasets. The results indicate that \textsf{ProMax} substantially improves base model performance and outperforms existing LLM-based recommendation approaches.
\end{itemize}

\begin{figure*}[t]
\setlength{\abovecaptionskip}{0.1cm}
\setlength{\belowcaptionskip}{0.1cm} 
  \centering
  \includegraphics[width=\linewidth]{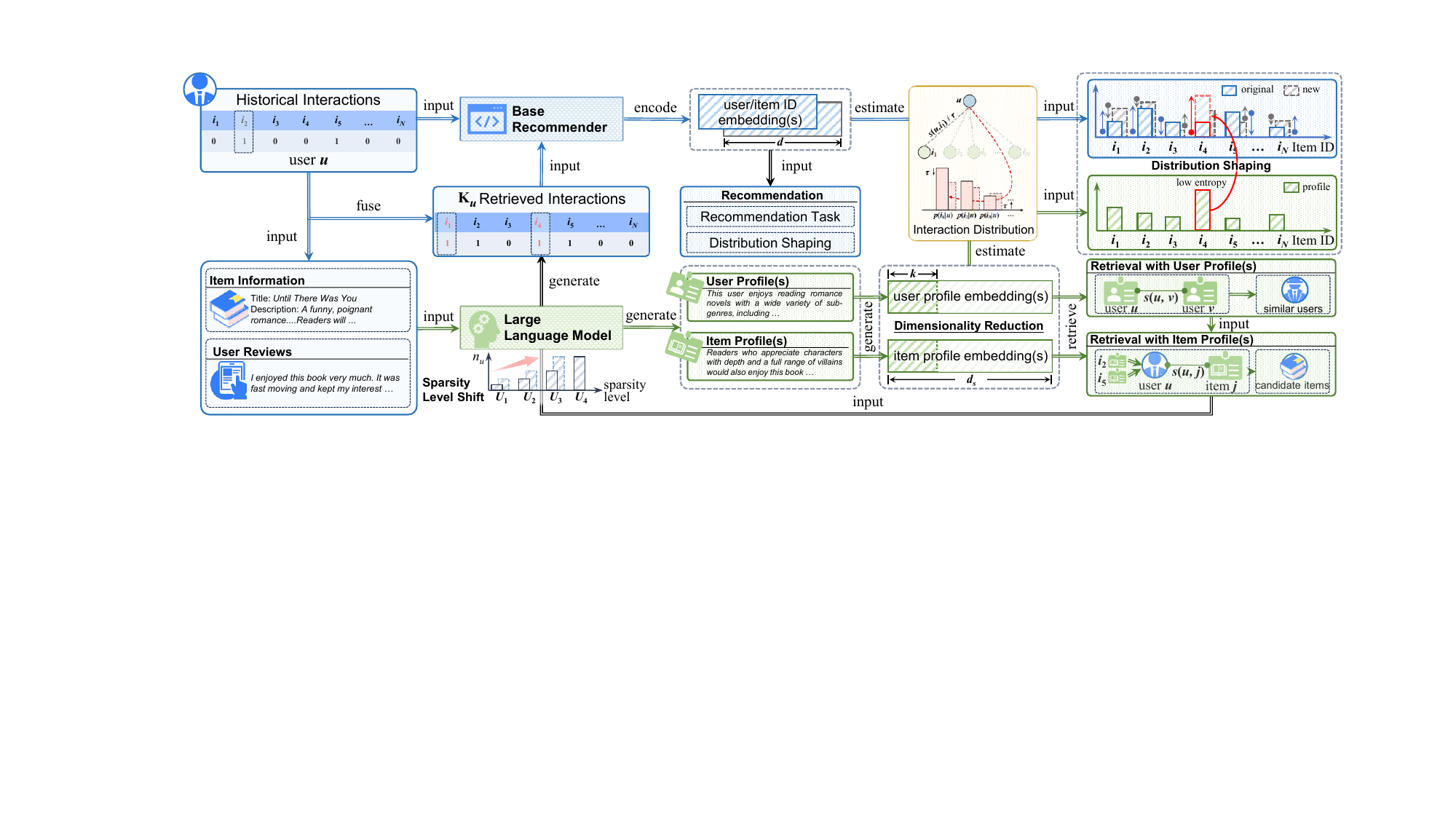}
  \caption{The information flow of the proposed \textsf{ProMax}.}
\label{fig_model}
\end{figure*}

\section{Methodology}
\subsection{Problem Formulation}
Within the general ID-based recommendation, we model the system as comprising $M$ users ($\mathcal U=\{u_1, u_2, ..., u_M\}$) and $N$ items ($\mathcal I=\{i_1, i_2, ..., i_N\}$), along with the interactions that capture how users engage with items \cite{he2017neural, he2020lightgcn}. For each user $u \in \mathcal U$, the historical interactions are represented by an interaction vector $\mathbf x_u \in \mathbb R^{1 \times N}$. If user $u$ has interacted with item $i$, the corresponding value in the vector $x_{ui}$ is set to 1.0. The objective of the recommender system is to learn a prediction model $f_\Theta$ that estimates the preference score $p_\Theta(j|u)$ for each unobserved item $j \in \mathcal I \setminus    \{i\}$ based on user and item embeddings $\mathbf z_u \in \mathbb R^{1 \times d}$ and $\mathbf z_j \in \mathbb R^{1 \times d}$, where $d$ is the embedding size \cite{rendle2009bpr}. Based on this, we propose the model-agnostic recommendation framework \textsf{ProMax}, as illustrated in Fig. \ref{fig_model}.

\subsection{Profiling with LLM}
\label{profile_modeling}
The strength of large language models in analyzing and generating text has been broadly recognized, a property that proves equally valuable in recommendation tasks \cite{wu2024survey}. As mentioned above, one particularly versatile method involves modeling users and items through fine-grained descriptions of their attributes, behavioral patterns, and contextual information, thereby explicitly representing true user preferences \cite{xi2024towards}. Without loss of generality, let $f_{\text{LLM}}$ denote an LLM. By providing item attributes such as titles, descriptions, and sampled user reviews as input, it is possible to derive an item-side profile $\mathcal P_i$ that captures the types of users likely to be interested in item $i$. Building on this, a user-side profile $\mathcal{P}_u$ can be constructed by aggregating item-to-user relationships, thereby characterizing the types of items that user $u$ tends to prefer \cite{ren2024representation}.

An important observation is that, although text-based profiles provide an explicit reflection of user preferences, they cannot be seamlessly utilized within recommendation model $f_\Theta$, thereby necessitating further transformation into structured representations. Therefore, a text embedding model $f_{\text{text}}$ is employed to transform all users' profiles into fixed-dimensional semantic representations:
\begin{equation}
\label{trans}
\mathbf C_{\mathcal U} = \{\mathbf c_{1}, \mathbf c_{2}, ..., \mathbf c_{M}\}, \ \text{where} \ \mathbf c_{u} = f_{\text{text}}(\mathcal P_{u}) \in \mathbb R^{1\times d_s}. 
\end{equation}
Here, $d_s$ denotes the dimensionality of the semantic representations obtained from the profiles. An analogous formulation can be established for the item side $\mathbf C_{\mathcal I}$. Typically, semantic representations $\mathbf c_u$ and $\mathbf c_i$  possess an intrinsic high-dimensional nature (\textit{i.e.}, $d_s \gg d$). As numerous studies have shown that high-dimensional semantic representations often include redundant dimensions with weak semantic relevance \cite{hu2025alphafuse}, we aim to enhance downstream recommendation performance through direct dimensionality reduction:

\begin{equation}
\label{mu}
\boldsymbol\mu = \frac{1}{M}\sum_{u \in \mathcal U} \mathbf c_u=\left (\frac{1}{M}\sum_{u \in \mathcal U}  c_{u,1},...,\frac{1}{M}\sum_{u \in \mathcal U}  c_{u,d_s} \right ) ,
\end{equation}

\begin{equation}
\label{cov}
\boldsymbol\Sigma = \frac{1}{M-1}\sum_{u \in \mathcal U} (\mathbf c_u - \boldsymbol\mu)(\mathbf c_u - \boldsymbol\mu)^\top=\mathbf V\text{diag}(\mathbf \Lambda )\mathbf V^\top ,
\end{equation}
where $\mathbf V=\{v_1, v_2, ..., v_{d_s}\}$ denotes the matrix of eigenvectors corresponding to the principal component directions, and $\mathbf \Lambda=\{\lambda_1, \lambda_2, ..., \lambda_{d_s}\}$ represents the eigenvalues. Subsequently, we select the top $\kappa$ eigenvectors  $\{v_1, v_2, ..., v_{\kappa}\}$ ($\kappa \ll d_s$) to balance semantic preservation with dimensionality reduction. Each user vector is first mean-centered and then projected onto these eigenvectors via inner products, yielding the transformed representation $\tilde{\mathbf c}_u$ \cite{hu2025alphafuse}:

\begin{equation}
\label{pca}
\tilde{\mathbf c}_u  = (v_1^\top(\mathbf c_u - \boldsymbol\mu), v_\kappa ^\top(\mathbf c_u - \boldsymbol\mu), ..., v_\kappa ^\top(\mathbf c_u - \boldsymbol\mu)) \in \mathbb R^{1\times \kappa}.
\end{equation}
At this stage, each semantic representation $\mathbf c_u$ is reduced to a low-dimensional form $\tilde{\mathbf c}_u$, which facilitates efficient processing of downstream tasks and focuses on semantically informative dimensions. A consistent definition is applied to the item side $\tilde{\mathbf c}_i$ as well.

\begin{figure}
\setlength{\abovecaptionskip}{0.0cm}
\setlength{\belowcaptionskip}{0.0cm} 
\centering
\subfigure[Profile Distribution]{\includegraphics[width=1.62in]{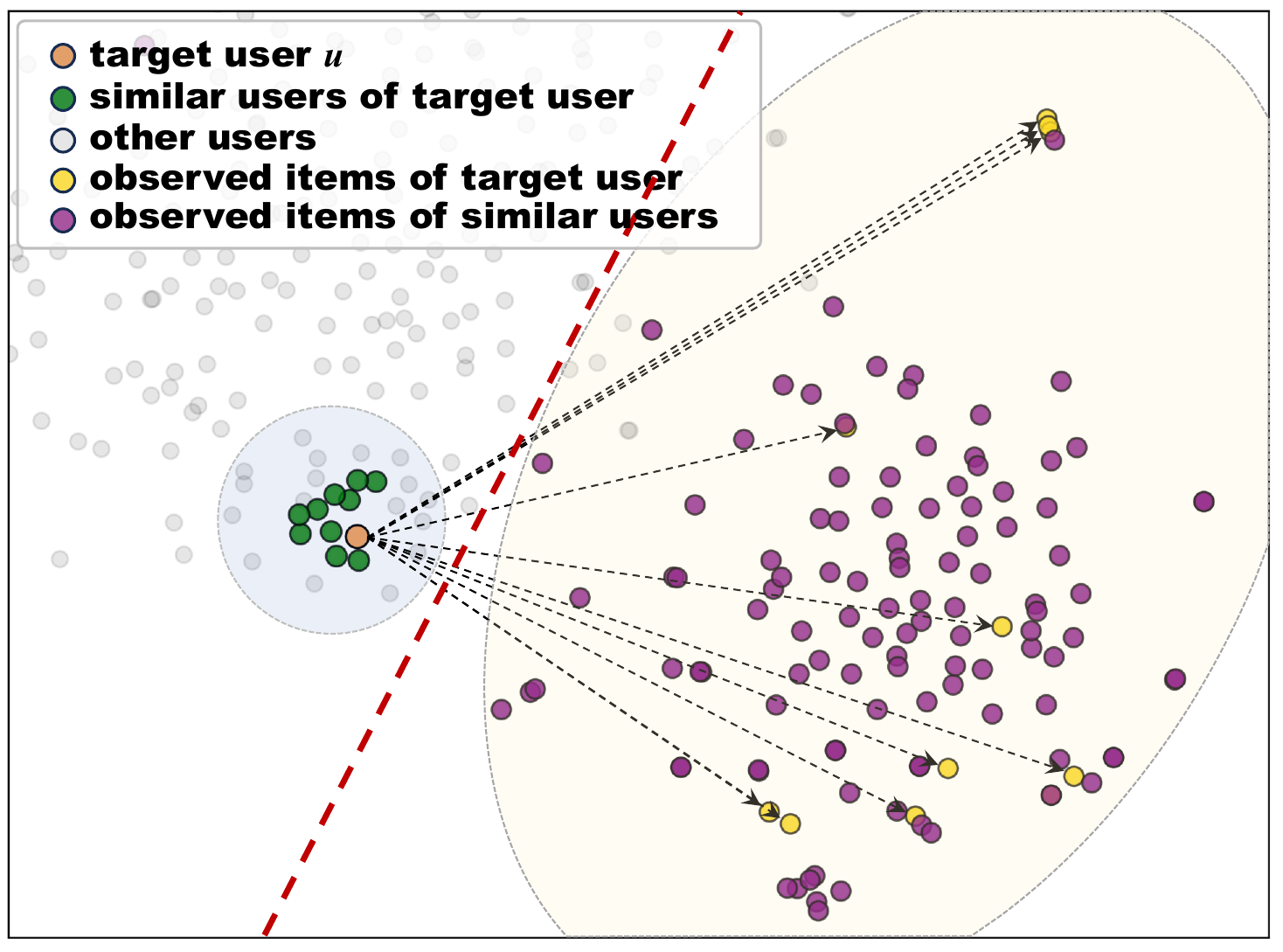}
\label{fig_similarity}}
\hfil
\subfigure[Sparsity Level Shift]{\includegraphics[width=1.62in]{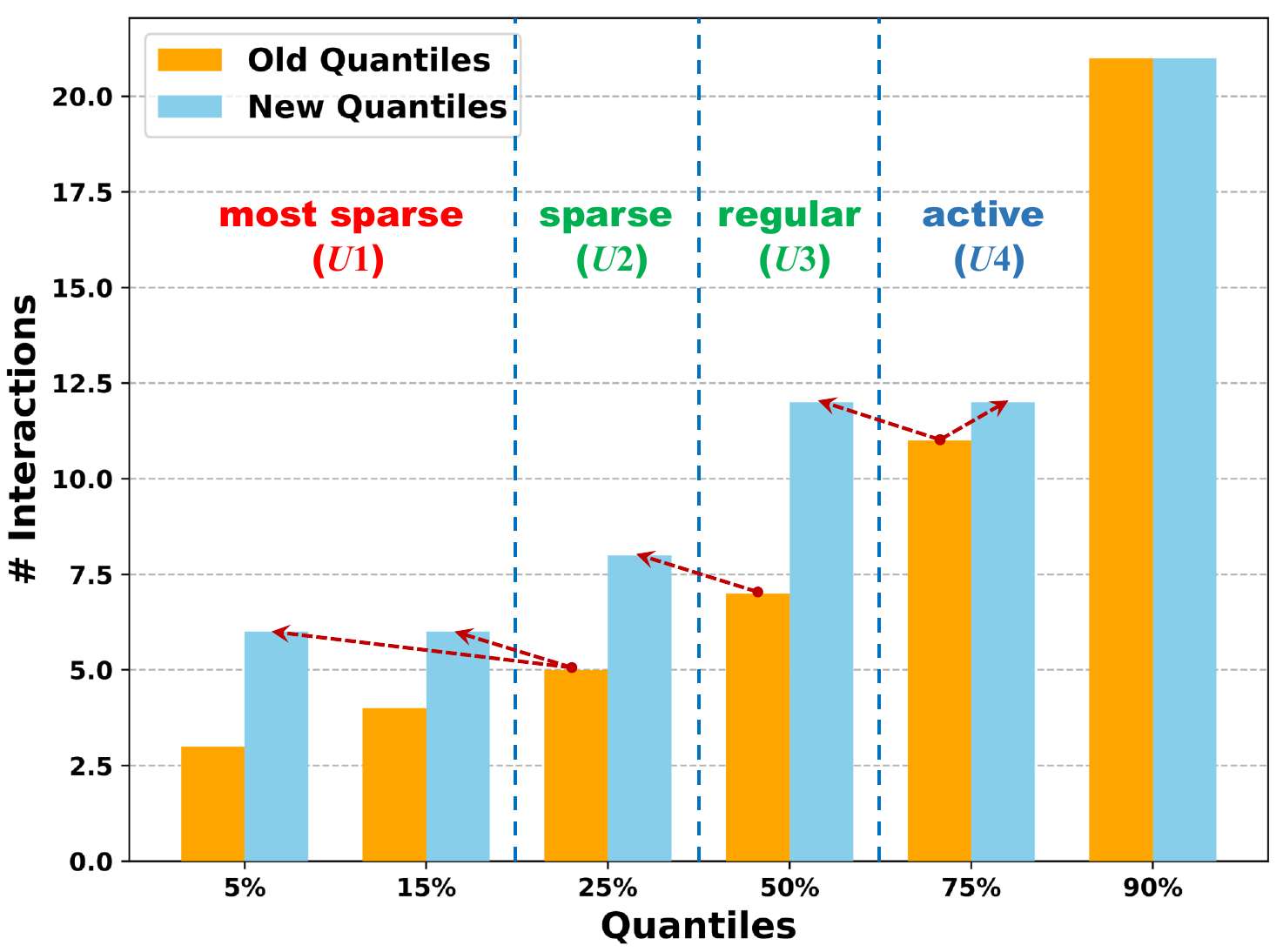}
\label{fig_quartile}}

\caption{(a) Distribution of profile representations for a sampled target user, along with their similar users and interacted items; (b) Comparison of quantile distributions before and after applying sparsity level shift on Amazon-Book dataset.}
\label{fig_motivation_rag}
\end{figure}

\subsection{Profile-Guided Candidate Retrieval}
\label{profile_retrieval}
We now have access to user and item profiles enriched with rich semantic information. Therefore, before directly leveraging these valuable additional resources to enhance recommendation, our primary objective is to examine whether these profiles are beneficial for the recommendation task, and in what ways they are beneficial.

\subsubsection{\textbf{Retrieval with User Profile(s)}} The core principle of CF is the assumption that users with similar behavioral patterns are likely to share consistent item preferences \cite{ren2024representation, he2017neural}. As an illustration, if two users engage with several overlapping items, they can be treated as similar, and their interaction records can be shared to support more accurate recommendations \cite{wang2019neural}. Drawing inspiration from dense retrieval \cite{fang2024scaling}, we can directly search for the set of the $\text{K}_1$ most similar users given a target user $u$: 

\begin{equation}
\label{simi_user}
\mathcal{R}_u^{\text{user}} \;=\; \operatorname{TopK} \left( \left\{s(\tilde{\mathbf{c}}_u, \tilde{\mathbf{c}}_v)|v \in \mathcal{U}\setminus\{u\}\right\} \right),
\end{equation}
where $s(\cdot, \cdot)$ denotes the cosine similarity function, $\mathbf c_u$ and $\mathbf c_v$ denote the profile representations corresponding to users $u$ and $v$, respectively. As shown in Fig. \ref{fig_similarity}, we visualize on Amazon-Book dataset \cite{ren2024representation} a randomly chosen user $u$ (orange dot), along with their similar users (green dots) and interacted items. The $\text K$ most similar users are those located closest to user $u$, and their observed distribution aligns well with common expectations.

\subsubsection{\textbf{Retrieval with Item Profile(s)}}
We then turn our focus to examining the distribution of items that these users have interacted with. Let $\mathcal M_u$ denote the set of items interacted with by user $u$.
Beyond this, we count the observed items $\mathcal M_v$ for each user $v$ in the similar user set $\mathcal{R}_u^{\text{user}}$, along with the corresponding number of interactions:
\begin{equation}
\label{simi_user_items}
\mathcal C_u = \{(i, \text{count}_{\mathcal R_u^{\text{user}}}(i)|i\in \mathcal M_v, i\notin \mathcal M_u, v\in \mathcal{R}_u^{\text{user}} )\},
\end{equation}
where $\mathcal M_u$ is the interacted item set of user $u$, and $\text{count}_{\mathcal R_u^{\text{user}}}(i)=\sum_{v\in\mathcal R_u^{\text{user}}}\mathbb I(i\in \mathcal M_v \wedge i\notin \mathcal M_u) $. It is worth noting that the statistics are restricted to the similar user set $\mathcal{R}_u^{\text{user}}$ of user $u$, thereby avoiding global bias \cite{chen2023bias}. Revisiting Fig. \ref{fig_similarity}, interacted item set $\mathcal M_u$ and set $\mathcal C_u$ are represented by yellow and purple dots, respectively, from which we derive the following observations:

\begin{itemize}[leftmargin=*]
\item[$\bullet$] The LLM-derived profile representations demonstrate favorable collaborative connectivity, as the items interacted with by user $u$ and their similar users show high similarity in the feature space.
\item[$\bullet$] As shown by the red dashed line in Fig. \ref{fig_similarity}, user and item profile representations are clearly separated in the feature space. Such separation is expected, given that their profiles are generated through different pipelines and carry different semantic signals.
\end{itemize}

The above observations indicate that the core advantage of profiles in enhancing recommendation stems from the inherent collaborative signals \cite{wang2019neural} they contain, which motivates the design of mechanisms to effectively connect user and item pipelines. Considering that the core of recommendation lies in suggesting unobserved items for the target user $u$ \cite{rendle2009bpr, he2017neural}, and that user and item profile semantics are separated in the feature space, we focus on performing dense retrieval over item-side profiles to explicitly capture the potential interactions of user $u$, following a design philosophy analogous to item-based CF models \cite{kabbur2013fism,zhang2023revisiting}:

\begin{equation}
\label{simi_item_candidate}
\mathcal R_u^{\text{item}}
= \operatorname{Rank}
\left( \left \{\frac{1}{\sqrt{\operatorname{count}_{\mathcal R_u^{\text{user}}}(i)}} s(\tilde{\mathbf c}_u^*,\tilde{\mathbf c}_i)\middle|i \in \mathcal C_u\right \} \right),
\end{equation}
where $\operatorname{Rank}(\cdot)$ sorts items in descending order. And we adopt mean pooling $\tilde{\mathbf c}_u^*=\sum_{j \in \mathcal M_u}\tilde{\mathbf c}_j/|\mathcal M_u|$ over the set of interacted items $\mathcal M_u$ to represent user $u$, enabling a finer-grained matching with the most similar items in $\mathcal{C}_u$ while mitigating the adverse effects of popularity bias from $\text{count}_{\mathcal R_u^{\text{user}}}$.

\subsubsection{\textbf{Sparsity Level Shift}}
The current challenge lies in determining how to appropriately control the number of items retrieved from the candidate set $\mathcal C_u$. If all retrieved items from $\mathcal R_u^{\text{item}}$ are appended to user $u$'s interactions, the original distribution may be severely distorted and excessive noise introduced. Conversely, when an equal number of new items is added across all users, the fixed quantity may disproportionately dominate the profiles of long-tail users, distorting their representations, while exerting only marginal influence on head users. Based on this, we propose \textit{sparsity level shift}, which constrains the number of retrieved items $\text{K}_{u}$ through a per-user allocation budget controlled by sparsity level:
\begin{equation}
\label{add_inter}
\text{K}_{u}=\text{max}(0, T(|\mathcal M_u|) - |\mathcal M_u| + 1).
\end{equation}
The above definition specifies the number of additional interactions $\text{K}_{u}$ required for user $u$. Accordingly, we define $\mathcal Q_\alpha$ denotes the $\alpha$-quantile of the interaction distribution across all users, \textit{i.e.}, $\mathcal Q_\alpha = \mathrm{Quantile}_\alpha\{|\mathcal M_u| : u \in \mathcal{U}\}$. For each user $u$, we identify the smallest quantile $ \mathcal Q_\alpha$ that is larger than the current interaction count $|\mathcal M_u|$, denoted as $T(|\mathcal M_u|)$:
\begin{equation}
\label{add_T}
T(|\mathcal M_u|)=\text{min}\{\mathcal Q \in \{\mathcal Q_{{0.25}},\mathcal  Q_{{0.5}},\mathcal  Q_{{0.75}}\}:|\mathcal M_u| < \mathcal Q\},
\end{equation}
where $\{\mathcal Q_{{0.25}},\mathcal  Q_{{0.5}},\mathcal  Q_{{0.75}}\}$ represents the first, the median, and the third quartile, thereby partitioning the entire population into four groups: most sparse ($U1$), sparse ($U2$), regular ($U3$), and active ($U4$).
At this point, we turn back to Eq. \ref{add_inter}. The adjustment $\text{K}_u$ is then computed to ensure that the user $u$’s interaction number reaches at least the next sparsity level. As shown in Fig. \ref{fig_quartile}, with sparsity level shift, the interaction counts of all user groups are elevated to the next tier. The effect is especially notable for mid-range users ($U2$ and $U3$), whereas active users ($U4$) are scarcely affected, given their already abundant interaction histories.

\begin{figure}[t]
\setlength{\abovecaptionskip}{0.1cm}
\setlength{\belowcaptionskip}{0.1cm} 
  \centering
  \includegraphics[width=\linewidth]{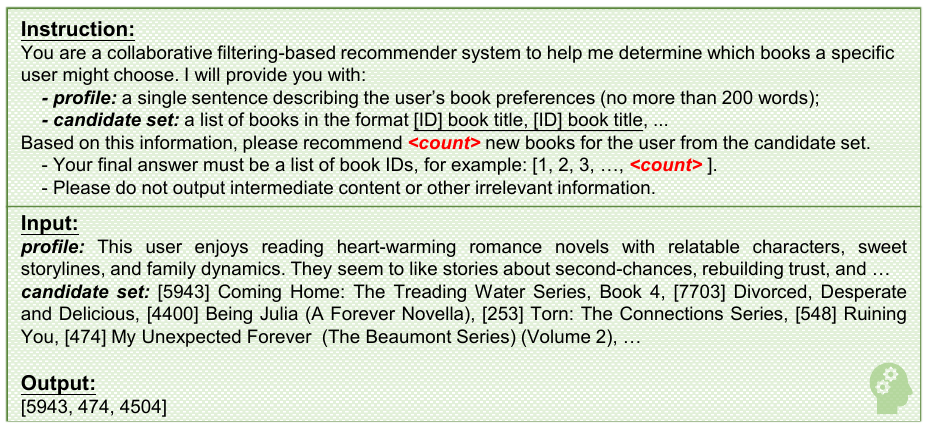}
  \caption{Detailed instruction for LLM-based re-ranking process, along with input and output examples.}
\label{fig_rag}
\end{figure}

After obtaining a per-user allocation budge $\text{K}_u$, we can directly apply it to $C_u$ and rewrite Eq. \ref{simi_item_candidate}. Considering the coarse-grained nature of dense retrieval, we further employ the LLM $f_{\text{LLM}}$ to re-rank all candidate item from $\mathcal R_u^{\text{item}}$ so as to better capture nuanced semantics and align more precisely with the user profile $\mathcal P_u$:

\begin{equation}
\label{simi_user_candidate}
\mathcal R_u^{\text{LLM}} =
f_{\text{LLM}}(\pi,\mathcal P_u, R_u^{\text{item}},\text K_u),
\end{equation}
where $\pi$ denotes the instruction component of the LLM prompt, and the detailed pipeline is shown in Fig. \ref{fig_rag}. Compared with directly selecting the top-$\text{K}_u$ highest-scoring items from the item set $\mathcal R_u^{\text{item}}$, LLM-based re-ranking is not only more flexible but also enables finer-grained filtering of candidate items, thereby mitigating exposure bias. On the other hand, using item set $\mathcal R_u^{\text{item}}$ as the candidate pool effectively prevents recommending non-existent items due to hallucination issue \cite{ji2023survey}. In the final step, we fuse the original interactions $\mathcal M_u$ with the retrieved interactions $\mathcal R_u^{\text{LLM}}$ to obtain an updated interaction set for user $u$: $\mathcal M'_u = \{i|i \in \mathcal M_u \, \text{or} \, i\in \mathcal R_u^{\text{LLM}} \}$. 

\subsection{Distribution Shaping}
Enhancing recommendation with semantic representations has been widely studied, including approaches that introduce adapters for dimensional alignment \cite{ren2024representation, hu2025alphafuse} or directly replace (or augment) ID embeddings in recommendation model $f_\Theta$ with semantic representations \cite{sheng2025language}. Nevertheless, these pointwise semantic alignment or enhancement processes not only increases complexity (\textit{e.g.}, by requiring additional MLP layers), but also fails to fully exploit the semantic information embedded in the profiles. Recalling the strong collaborative connectivity illustrated in Fig. \ref{fig_similarity}, the strong collaborative connectivity can be viewed as a recommendation indicator informed by semantic priors. This indicator uncovers latent relationships between user and item profiles while simultaneously offering an auxiliary supervisory signal to base recommendation model $f_\Theta$. To provide direct guidance for recommendation while avoiding extra processes, we start by defining the user $u$'s distribution over the entire item set:
\begin{equation}
\label{distribution}
p_{*}(i|u) = \text{Softmax}_{i\in\mathcal I}(s(\mathbf e_u, \mathbf e_i)/\tau), \quad \forall i \in \mathcal{I},
\end{equation}
where notation “$*$” is used to indicate the source of the user and item representations, $\mathbf e_u$ and $\mathbf e_i$  (for $f_\Theta$, we have $\mathbf e=\mathbf z$, whereas for $f_\text{LLM}$ we have $\mathbf e=\tilde{\mathbf c}$). The temperature coefficient $\tau$ controls the sharpness of the probability distribution \cite{chen2023adap}. Based on this, our ultimate objective is to shape the conditional distribution $p_{\Theta}(i|u)$ produced by the recommendation model $f_\Theta$ such that it more accurately reflects the intrinsic preferences of user $u$. 

\subsubsection{\textbf{Supervised Distribution Reshaping}}
As previously noted, the LLM-driven profiles act as recommendation indicators, and the corresponding user–item distributions $p_{\text{LLM}}(i|u)$ are employed to guide the model $f_\Theta$ to transcend simple memorization of observed interactions $\mathcal M_u$ and instead produce preference distributions over unobserved items:
\begin{equation}
\label{supervised_align}
\mathcal L_{\text{SDR}}=-\frac{1}{|\mathcal U_\mathcal B|} \sum_{u\in \mathcal U_\mathcal B}w(u)\cdot\sum_{i\in\mathcal I}\text{sg}[p_{\text{LLM}}(i|u)]\cdot\text{log}p_{\Theta}(i|u),
\end{equation}
where $\text{sg}[\cdot]$ represents the stop-gradient operator, $\mathcal U_\mathcal B$ denotes the set of users contained in a mini-batch $\mathcal B$, and $w(u)$ is the uncertainty-based weight of user $u$. In this process, the LLM-derived profile representations $\{\tilde{\mathbf c}_u, \tilde{\mathbf c}_i\}$ are not directly injected into the training of the recommendation model $f_\Theta$. This design obviates the need for additional dimensional or semantic alignment modules, and instead provides a principled yet lightweight way of guiding the recommendation model $f_\Theta$ through the interaction distribution $p_{\text{LLM}}(i|u)$. Since these representations $\{\tilde{\mathbf c}_u, \tilde{\mathbf c}_i\}$ are not actually involved in training, and are reduced in dimensionality following the approach introduced in Section \ref{profile_modeling}, the distribution $p_{\text{LLM}}(i|u)$ can be pre-computed offline or computed on the fly within each batch $\mathcal{B}$, while introducing negligible additional time complexity.

Given the prevalence of long-tail items, it becomes crucial to investigate whether the profile distribution $p_{\text{LLM}}(i|u)$ demonstrates dispersion in its probabilities, as such dispersion reflects the LLM’s uncertainty regarding the extent to which a user may favor these items. To this end, we introduce an uncertainty weighting $w(u)$ in Eq. \ref{supervised_align} to quantify the confidence of the profile distribution for each user $u$. More specifically, we introduce the entropy of the interaction distribution as induced by the user/item profiles \cite{shannon2001mathematical}:
\begin{equation}
\label{entropy}
H(u) = - \sum_{i\in\mathcal I}p_{\text{LLM}}(i|u) \cdot\text{log}p_{\text{LLM}}(i|u).
\end{equation}
Here, $H(p_{\text{LLM}}(i|u)) $ measures the uncertainty embedded in the distribution $p_{\text{LLM}}(i|u)$ by aggregating over all items $i \in \mathcal I$. A larger entropy value suggests that the LLM distributes probabilities more diffusely over items, signaling greater uncertainty regarding user $u$’s preferences, while a smaller entropy corresponds to more confident and concentrated predictions. Since the maximum entropy depends on the size of the item set $|\mathcal I|$, we constrain the uncertainty within [0,1] to obtain a unified measure: $\hat{H}(u) = H(u)/\text{log}|\mathcal I|$. Finally, we assign greater weight to the supervisory signal of the recommendation model $f_\Theta$ when the profile distribution $p_{\text{LLM}}(i|u) $ is more reliable (\textit{i.e.}, exhibits lower entropy $\hat{H}(u)$), formulated as $w(u) = 1 - \hat{H}(u)$.

\subsubsection{\textbf{Self-Supervised Distribution Reshaping}}
While the adaptive weight $w(u)$ is employed to calibrate the supervisory signal derived from the profile distribution $H(p_{\text{LLM}}(i|u))$, a critical concern arises from the inherent instability, noise, and hallucination problems in $f_{\text{LLM}}$ \cite{ji2023survey}. If the profile distribution is of poor quality from the outset, its capacity to provide effective guidance for the recommendation model $f_\Theta$ may be limited \cite{sobal2024mathbb}. 

Referring back to Section \ref{profile_retrieval}, we leverage these profiles to identify additional items $\mathcal R_u^{\text{LLM}}$ of potential interest to user $u$, thereby constructing an augmented set $\mathcal M'_u$ on top of the original interactions $\mathcal M_u$. For the same user $u$, different interaction sets $\mathcal M_u$ and $\mathcal M'_u$ could yield semantically distinct representations. Therefore, we enforce bidirectional consistency maximization for user $u$ between the two different interaction sets, thereby encouraging the recommendation model $f_\Theta$ to further exploit more collaborative signals from both perspectives:
\begin{equation}
\label{cross_align}
\mathcal L_{\text{S}^2\text{DR}} = -\frac{1}{2|\mathcal B|}\sum_{<u,i>\in \mathcal B}\sum_{o \in \{\mathcal M_u,\mathcal M'_u\}}\left [\text{log}p_\Theta(i_{\bar{o}}|u_{o}) \right ],
\end{equation}
where $u_o$ and $i_o$ denote the user and item semantic sources obtained by feeding the interaction set $o \in \{\mathcal M_u,\mathcal M'_u\}$ into the recommendation model $f_\Theta$, while $\bar{o}$ represents the opposite view of interaction set $o$. In conclusion, we further leverage the semantic information embedded in LLM-driven profiles and introduce a self-supervised mechanism that encourages the recommendation model to examine discrepancies across perspectives \cite{yu2023self}, thereby mitigating bias from excessive reliance on the original profiles. Building on this, we also introduce contrastive alignment processes between homogeneous nodes (users or items) to further enhance the model’s expressive capabilities \cite{zhang_lightccf_2025}.

{\small
\begin{algorithm}[t]
\caption{The training process of \textsf{ProMax}}
\label{algorithms}
\KwIn{user–item interaction $\mathbf X$, large language model $f_{\text{LLM}}$, text embedding model $f_{\text{text}}$, and base recommendation model $f_{\Theta}$;}
\begin{algorithmic}[1]
\STATE //\texttt{\textbf{pre-processing stage via} $f_{\text{LLM}}$}
\STATE generate all item profile $\{\mathcal P_i\}$ by $f_{\text{LLM}}$ and transform them into $\kappa$-dimensional representation by $f_\text{text}$ (Eq. \ref{pca});
\FOR{$u\in \mathcal U$}
\STATE generate a user profile $\mathcal P_u$ by $f_{\text{LLM}}$ and transform it into $\kappa$-dimensional representation by $f_\text{text}$ (Eq. \ref{pca});
\STATE retrieve a similar user set $\mathcal R_u^{\text{user}}$ and an item set $\mathcal R_u^{\text{item}}$ by Eqs. \ref{simi_user}, \ref{simi_item_candidate};
\STATE retrieve a new item set $\mathcal R_u^{\text{LLM}}$ from $\mathcal R_u^{\text{item}}$ by $f_\text{LLM}$ (Eq. \ref{simi_user_candidate});
\STATE update the interaction set $\mathcal M'_u = \{i|i \in \mathcal M_u \, \text{or} \, i\in \mathcal R_u^{\text{LLM}} \}$ with $\text{K}_u$ (Eq. \ref{add_inter});
\ENDFOR
\STATE \vspace{0.0em}
\STATE //\texttt{\textbf{training stage for recommendation model}} $f_\Theta$
\STATE initialize model parameters for $f_{\Theta}$;
\WHILE {\textsf{ProMax} has not converge}
\STATE sample a mini-batch of user-item pairs $\mathcal B$ from $\mathbf X$;
\FOR{$<u,i>\in \mathcal B$}
\STATE perform forward encoding $f_\Theta$;
\STATE  construct user distribution $p_{\Theta}(i|u)$ and $p_{\text{LLM}}(i|u)$ by Eq. \ref{distribution};
\STATE  construct $p_{\Theta}(i_{\mathcal M_u'}|u_{\mathcal M_u})$ and $p_{\Theta}(i_{\mathcal M_u}|u_{\mathcal M_u'})$ by Eq. \ref{distribution};
\STATE  calculate two distribution reshaping losses $\mathcal L_{\text{SDR}}$ by Eq. \ref{supervised_align} and  $\mathcal L_{\text{S}^2\text{DR}}$ by Eq. \ref{cross_align};
\STATE  calculate the recommendation loss $\mathcal L_\text{rec}$;
\STATE calculate the total loss $\mathcal L_{\textsf{ProMax}} = \mathcal L_{\text{rec}} + \lambda_1 \mathcal L_{\text{SDR}} + \lambda_2 \mathcal L_{\text{S}^2\text{DR}}$;
\ENDFOR
\STATE average gradients from mini-batch;
\STATE update model parameters by descending the gradients $\nabla_{\Theta}\mathcal L_{\textsf{ProMax}}$;
\ENDWHILE
\RETURN model parameters $\Theta$;
\end{algorithmic}
\end{algorithm}
}

Finally, let $\mathcal L_{\text{rec}}$ denote the recommendation loss of the base recommender $f_\Theta$ (\textit{e.g.}, pairwise ranking loss \cite{rendle2009bpr}). The complete training objective of \textsf{ProMax} is then defined as:
\begin{equation}
\label{loss}
\mathcal L_{\textsf{ProMax}} = \mathcal L_{\text{rec}} + \lambda_1 \cdot \mathcal L_{\text{SDR}} + \lambda_2 \cdot \mathcal L_{\text{S}^2\text{DR}},
\end{equation}
where $\lambda_1$ and $\lambda_2$ are adjustable weights introduced to ensure that the losses are on the same scale. In conclusion, \textsf{ProMax} is a plug-and-play recommendation framework where LLM-driven profiles are excluded from direct model training, and retrieval and distribution computations can be efficiently pre-processed. 

\subsection{Model Analysis}
The complete training process of \textsf{ProMax} is presented in Algorithm \ref{algorithms}. First, we construct profiles for all users and items and map them into $k$-dimensional representations to better support downstream tasks (lines 2-4). To fully exploit the potential of these profiles, we perform dense retrieval processes on both the user and item sides to capture potential user-item interactions (line 5). Subsequently, to obtain finer-grained semantic information, we employ the $f_\text{LLM}$ again to re-filter the candidate item set $\mathcal R_u^{\text{item}}$ (line 6), and by controlling the sparsity shift, we select top-$\text{K}_u$ new items for user $u$ to update the interaction set $\mathcal M_u'$ (line 7). The profile representations and the updated interaction set are utilized for training the downstream recommendation model $f_\Theta$ (lines 15-17). Two additional loss terms, $\mathcal L_{\text{SDR}}$ and $\mathcal L_{\text{S}^2\text{DR}}$ (line 18), are computed on top of the original recommendation loss $\mathcal L_{\text{rec}}$ (line 19).

Since the pre-processing stage is independent of the recommendation model training, it can be completed in advance without incurring additional training overhead. Moreover, since the dimensionality of the profile representations has been significantly reduced (\textit{i.e.,} $\kappa \ll d_s$), , the time required for the dense retrieval processes is negligible in practice. During the training stage for recommendation model $f_\Theta$, \textsf{ProMax} serves as a plug-and-play framework that does not alter the training procedure of the base model $f_\Theta$, and only computes two additional loss terms, $\mathcal L_{\text{SDR}}$ and $\mathcal L_{\text{S}^2\text{DR}}$. The time complexity of the former is $O(|\mathcal B|N\text{max}(d, \kappa))$, while that of the latter is $O(|\mathcal B|^2d)$, where $|\mathcal B|$ is the batch size, $N$ is the number of items, $d$ is the embedding dimension, and $\kappa$ is the dimension of the profile representation. The time complexities of these two losses are both lower than that of the base model $f_\Theta$’s encoding, so they do not significantly increase the overall training time of the recommendation model $f_\Theta$. Details can be found in Section \ref{section_time}.


\begin{table}[t]
\small
\setlength{\abovecaptionskip}{0.1cm}
\setlength{\belowcaptionskip}{0.1cm} 
  \caption{ Statistics of the datasets.}
  \label{dataset}
  \begin{tabular}{l|c|c|c|c}
    \hline
    \textbf{Dataset}&\textbf{\#Users}&\textbf{\#Items}&\textbf{\#Interactions}&\textbf{Sparsity}\\
    \hline
    \hline
    \textbf{Amazon-Book}&11,000&9,332&200,860&99.80\%\\
    \textbf{Yelp}&11,091&11,010&277,535&99.77\%\\
    \textbf{Steam}&23,310&5,237&525,922&99.57\%\\
    \hline
  \end{tabular}
\end{table}

\begin{table*}
\setlength{\abovecaptionskip}{0.1cm}
\setlength{\belowcaptionskip}{0.1cm} 
  \caption{Overall performance comparisons on Amazon-Book, Yelp, and Steam datasets \textit{w.r.t.} Recall@N (abbreviated as R@N) and NDCG@N (abbreviated as N@N), where $\text{N}\in [10, 20]$. The best-performing model on each dataset and metric is highlighted in \textbf{bolded} ‘Improv.\%’ indicates the relative improvement of \textsf{ProMax} over the best baseline based on a two-tailed paired t-test. }
  \label{performance1}
  \begin{tabular}{l|cccc|cccc|cccc}
    \hline
    &\multicolumn{4}{c|}{\textbf{Amazon-Book}}&\multicolumn{4}{c|}{\textbf{Yelp}}&\multicolumn{4}{c}{\textbf{Steam}}\\
	\cline{2-13}		
	&R@10&R@20&N@10&N@20&R@10&R@20&N@10&N@20&R@10&R@20&N@10&N@20\\
    \hline
    \hline
    Base             &0.0915 &0.1411 &0.0694 &0.0856 &0.0706 &0.1157 &0.0580 &0.0733 &0.0852 &0.1348 &0.0687 &0.0855\\
    \hline
	KAR \cite{xi2024towards}             &0.0934 &0.1416 &0.0705 &0.0860 &0.0740 &0.1194 &0.0602 &0.0756 &0.0854&0.1353&0.0690&0.0854\\
	LLMRec \cite{wei2024llmrec}             &0.0963 &0.1469 &0.0715 &0.0855 &0.0744 &0.1203 &0.0605 &0.0751 &0.0901&0.1431&0.0722&0.0901\\

    CARec \cite{wang2024collaborative}             &0.0931 &0.1392 &0.0702 &0.0854 &0.0700 &0.1130 &0.0568 &0.0714 &0.0900 &0.1384 &0.0713 &0.0880\\

    RLMRec-C \cite{ren2024representation}     &0.0969 &0.1483 &0.0734 &0.0903 &0.0754 &0.1230 &0.0614 &0.0776 &0.0895&0.1421&0.0724&0.0902\\
	RLMRec-G \cite{ren2024representation}          &0.0948 &0.1446 &0.0724 &0.0887 &0.0734 &0.1209 &0.0600 &0.0761 &0.0907&0.1433&0.0729&0.0907\\

	AlphaRec \cite{sheng2025language}      &0.0941 &0.1412 &0.0721 &0.0873 &0.0726 &0.1212 &0.0586 &0.0751 &0.0897&0.1404&0.0718&0.0889\\
    
	AlphaRec+ \cite{sheng2025language}      &0.0945 &0.1421 &0.0710 &0.0835 &0.0732 &0.1213 &0.0590 &0.0752 &0.0900 &0.1420 &0.0719 &0.0898\\

    LLMESR \cite{liu2024llm}      &0.0947 &0.1435 &0.0722 &0.1255 &0.0739 &0.1207 &0.0599 &0.0758 &0.0784 &0.1281 &0.0631 &0.0798\\
    
    AlphaFuse \cite{hu2025alphafuse}      &0.0952 &0.1427 &0.0716 &0.0870 &0.0720 &0.1175 &0.0584 &0.0739& 0.0895 &0.1396 &0.0733 &0.0901\\
    IRLLRec \cite{wang2025intent}    &0.0996 &0.1513 &0.0759 &0.0928 &0.0776 &0.1267 &0.0632 &0.0798 &0.0918&0.1445&0.0737&0.0919\\
    ProEx \cite{zhang2025proex}    &0.1008 &0.1533 &0.0770 &0.0940 &0.0800 &0.1308 &0.0653 &0.0826 &0.0937&0.1473&0.0756&0.0939\\
	\hline
	\textbf{\textsf{ProMax} (Ours)}&\textbf{0.1079} &\textbf{0.1603} & \textbf{0.0835} & \textbf{0.1006} &\textbf{0.0826} &\textbf{0.1338} &\textbf{0.0684} &\textbf{0.0858} &\textbf{0.0965}&\textbf{0.1500} &\textbf{0.0781}&\textbf{0.0964}\\
    \hline
    \hline
    Improv.\%&7.04\%&4.57\%&8.44\%&7.02\%&3.25\%&2.29\%&4.75\%&3.87\%&2.99\%&1.83\%&3.31\%&2.66\%\\
	$p$-value&$3.6\text{e}^{-7}$&$2.2\text{e}^{-7}$&$9.6\text{e}^{-9}$&$8.0\text{e}^{-9}$&$1.2\text{e}^{-5}$&$9.3\text{e}^{-8}$&$4.3\text{e}^{-8}$&$5.9\text{e}^{-7}$&$1.3\text{e}^{-7}$&$3.3\text{e}^{-7}$&$1.9\text{e}^{-8}$&$2.1\text{e}^{-8}$\\
    \hline
  \end{tabular}
\end{table*}

\begin{figure*}[t]
\setlength{\abovecaptionskip}{0.1cm}
\setlength{\belowcaptionskip}{0.1cm} 
  \centering
  \includegraphics[width=\linewidth]{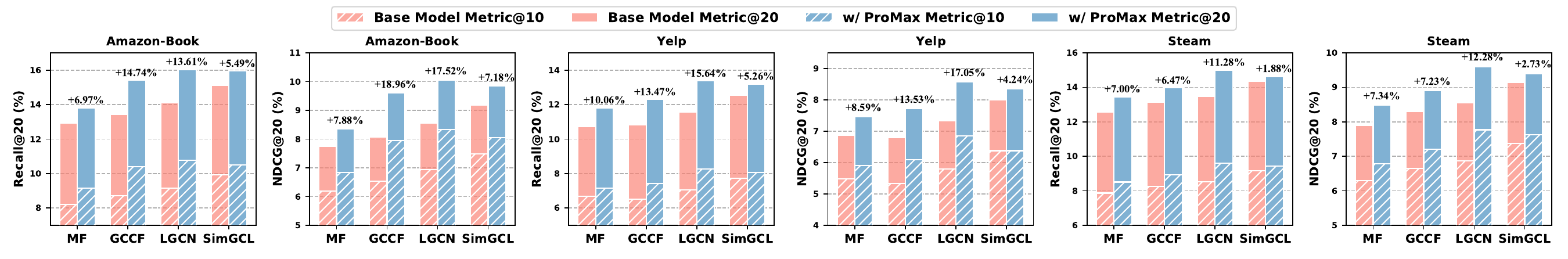}
  \caption{Performance comparison between \textsf{ProMax} and base models MF \cite{rendle2009bpr}, LR-GCCF (abbreviated as GCCF) \cite{chen2020revisiting}, LightGCN (abbreviated as LGCN) \cite{he2020lightgcn}, and SimGCL \cite{yu2022graph} on Amazon-Book, Yelp, and Steam datasets \textit{w.r.t.} Recall@10/20 and NDCG@10/20.}
\label{fig_improve}
\end{figure*}

\section{Experiments}
In this section, we present detailed experiments and analyses to demonstrate the effectiveness of \textsf{ProMax}.

\subsection{Experimental Settings}
\subsubsection{\textbf{Datasets}} We employ three extremely sparse datasets Amazon-Book, Yelp, and Steam \cite{ren2024representation} in our experiments, which are varied in scale, field, and
sparsity. Following prior studies \cite{ he2020lightgcn}, we focus on implicit feedback \cite{he2017neural} and discard interactions with ratings below 3.0 \cite{ren2024representation}.  Items that a user has interacted with are considered positive samples, and other items apart from that are seen as negative samples for that user. Each dataset is then partitioned into training, validation, and test sets using a 3:1:1 ratio. The statistical information of three datasets is shown in Table \ref{dataset}.

\subsubsection{\textbf{LLM-enhanced Baselines}}
To facilitate a comprehensive comparison, we adopt following LLM-enhanced user profiling methods as baselines:
\begin{itemize}[leftmargin=*]
    \item \textbf{KAR} \cite{xi2024towards} constructs textual profiles for users and items, and integrates the LLM-enhanced representations with base recommender models through a hybrid-expert adaptor.
    \item \textbf{LLMRec} \cite{wei2024llmrec} improves data reliability by combining LLM–based graph augmentation with a denoising mechanism for robust recommendation modeling.
    \item \textbf{CARec} \cite{wang2024collaborative} fuses collaborative filtering signals with the semantic representations produced by large language models through a collaborative alignment mechanism.
    \item \textbf{RLMRec} \cite{ren2024representation} first constructs user and item profiles, followed by aligning user and item semantic representations across the LLM-enhanced and recommendation representations.
    \item \textbf{AlphaRec} \cite{sheng2025language} performs non-linear mapping and multi-layer graph convolution directly on LLM-enhanced item representations for recommendation task.
    \item \textbf{LLMESR} \cite{liu2024llm} utilizes LLM-driven representations to enhance recommendation, and employs retrieval-augmented self-distillation to enhance user preference representations.
    \item \textbf{AlphaFuse} \cite{hu2025alphafuse} distinguishes LLM-driven embeddings into a semantic-rich row space and a semantic-sparse null space, and integrates ID embeddings into the latter.
    \item \textbf{IRLLRec} \cite{wang2025intent} not only leverages LLMs to construct initial user and item profiles, but also exploits them to construct multimodal intents and enhance recommendation.
    \item \textbf{ProEx} \cite{zhang2025proex} is a recently proposed LLM-enhanced method that leverages chain-of-thought reasoning to construct multiple profiles for users and items and extract semantic invariances.

\end{itemize}

For RLMRec \cite{ren2024representation}, we denote the variants with generative and contrastive strategies as \textbf{RLMRec-G} and \textbf{RLMRec-C}, respectively. Regarding AlphaRec \cite{sheng2025language}, which originally leverages only item-side semantic representations, we introduce a variant that additionally incorporates user-side representations, referred to as \textbf{AlphaRec+}. It is worth noting that LLMESR \cite{liu2024llm} and AlphaFuse \cite{hu2025alphafuse} are originally designed as sequential recommendation models (\textit{e.g.}, SASRec \cite{kang2018self}). In this work, we adapt their official implementations to the general recommendation setting for a consistent comparison.

We do not include fine-tuning-based methods (\textit{e.g.}, LLaRA \cite{liao2024llara} and SPRec \cite{gao2025sprec}) in our comparison experiments, since their objectives are orthogonal to ours. These methods aim to fine-tune LLMs for sequential recommendation \cite{sheng2025language}, whereas our study focuses on leveraging LLMs without fine-tuning in a general recommendation setting \cite{he2020lightgcn, ren2024representation}.

\subsubsection{\textbf{Implementation Details}} We implement \textsf{ProMax} by PyTorch\footnote{https://github.com/BlueGhostYi/ProRec}. To ensure fairness in comparison, all models are initialized using Xavier initialization \cite{glorot2010understanding} and optimized with the Adam optimizer \cite{kingma2014adam}. LightGCN \cite{he2020lightgcn} is chosen as the base model for all LLM-enhanced methods because of its simplicity and generality, which is configured with three graph convolutional layers and an embedding size of 32 by default, without any additional dropout mechanism. The learning rate and batch size are set to 1.0e-3 and 4,096, respectively. To ensure a fair comparison, the user and item profiles are derived from \cite{ren2024representation}. Based on this, we use OpenAI’s GPT-4o-mini to perform the re-ranking process (\textit{i.e.,} the process shown in Fig. \ref{fig_rag}). For \textsf{ProMax}, the original profile representations are compressed to match the embedding size (\textit{i.e.,} $\kappa=d=32$), and temperature coefficient $\tau$ is set to 0.2 for Eq. \ref{distribution} by default \cite{yu2022graph}. In the retrieval stage, we empirically set $\text{K}=100$ for Eq. \ref{simi_user} to reduce computational cost. The hyperparameters $\lambda_1$ and $\lambda_2$ constrained within the ranges $[0,2]$ and $[0,0.2]$, respectively, to balance the magnitude of different loss terms. We evaluate recommendation performance using Recall@N and NDCG@N (N$\in [10, 20]$) \cite{he2017neural}, and adopt a full-ranking evaluation strategy for each user \cite{yin2025device}. Early stopping is applied if Recall@20 on the validation set fails to improve for 20 consecutive epochs. All experiments are run with 10 random seeds, and the average results are reported.

\subsection{Performance Comparisons}
\subsubsection{\textbf{Performance Comparison with LLM-enhanced Methods}}

To verify the effectiveness of \textsf{ProMax}, we perform comparative experiments against other LLM-enhanced recommendation methods. Table \ref{performance1} reports the recommendation performance of \textsf{ProMax} along with all LLM-enhanced methods on the LightGCN backbone \cite{he2020lightgcn}, leading to the following observations:

 Intuitively, \textsf{ProMax} consistently achieves the best recommendation performance across all base models on three datasets. Quantitatively, integrating \textsf{ProMax} into LightGCN yields improvements of 8.44\%, 4.75\%, and 3.31\% over the best baseline (\textit{i.e.}, ProEx \cite{zhang2025proex}) \textit{w.r.t.} NDCG@10 on Amazon-Book, Yelp, and Steam datasets, respectively. The above results demonstrate the effectiveness of \textsf{ProMax}, which we attribute to the proposed supervised and self-supervised distribution reshaping process.
 
 While prior work such as RLMRec \cite{ren2024representation} leverages user and item profiles to enhance recommendation, the profile information is not fully utilized. IRLLRec \cite{wang2025intent} extends this line of research by introducing intent profiles, achieving further performance gains at the cost of more complex profile construction and training procedures. AlphaRec \cite{sheng2025language} attempts to directly replace ID embeddings with textual representations for training, which may hinder the recommender model from capturing sufficient collaborative signals. Moreover, the limited number of parameters makes the model prone to overfitting and suboptimal performance. Although ProEx \cite{zhang2025proex} achieves further improvements by multi-profiling, constructing additional profiles incurs extra costs, while the performance gains remain limited. In contrast, \textsf{ProMax} effectively unlocks the potential of a single profile, enabling better performance with substantially lower cost.
 
Finally, the recent works LLMESR \cite{liu2024llm} and AlphaFuse \cite{hu2025alphafuse}, which attempt to fuse ID and semantic representations, also fail to deliver satisfactory performance. On the one hand, their original architectures were designed for interaction sequence modeling, making them less suitable for pure CF tasks. On the other hand, the lack of a necessary alignment between the interaction space and the language space leads to semantic loss and introduces unnecessary noise during forced fusion.

\subsubsection{\textbf{Model-agnostic Performance Comparison}}
To evaluate the generalization capability of \textsf{ProMax}, we integrate it with four classic discriminative methods, namely MF \cite{rendle2009bpr}, LR-GCCF \cite{chen2020revisiting}, LightGCN \cite{he2020lightgcn}, and SimGCL \cite{yu2022graph}. The experimental results are reported in Fig. \ref{fig_improve}. It is evident that incorporating \textsf{ProMax} yields performance gains for all base models on every dataset. For example, when applied to LightGCN, \textsf{ProMax} improves Recall@20 by 13.3\%, 15.64\%, and 12.28\% on Amazon-Book, Yelp, and Steam, respectively, highlighting its strong generalization and effectiveness. The second base model that benefits the most is GCCF. This is because the original model is constrained by redundant feature transformation, whereas the additional supervision introduced by \textsf{ProMax} enables it to overcome this limitation and rediscover its performance potential. On the other hand, the improvement brought by \textsf{ProMax} to SimGCL is not particularly pronounced. A possible reason is that SimGCL already leverages self-supervised learning to provide abundant supervisory signals, so the integration of \textsf{ProMax} serves more as a complement rather than delivering a qualitative leap.

\subsubsection{\textbf{Performance Comparison w.r.t. Data Sparsity}}
\begin{figure}[t]
\setlength{\abovecaptionskip}{0.0cm}
\setlength{\belowcaptionskip}{0.0cm} 
  \centering
\subfigure[Amazon-Book: (left) Quantile distribution; (right) Recommendation performance]
{\includegraphics[width=\linewidth]{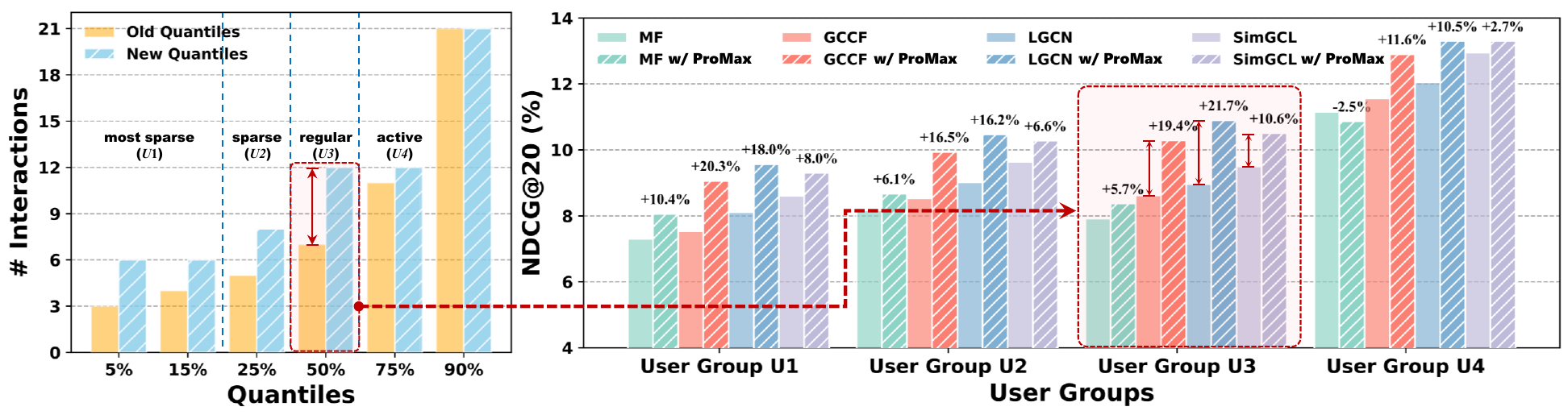}
\label{fig_amazon_sparsity}}
\hfil
\subfigure[Yelp: (left) Quantile distribution; (right) Recommendation performance]
{\includegraphics[width=\linewidth]{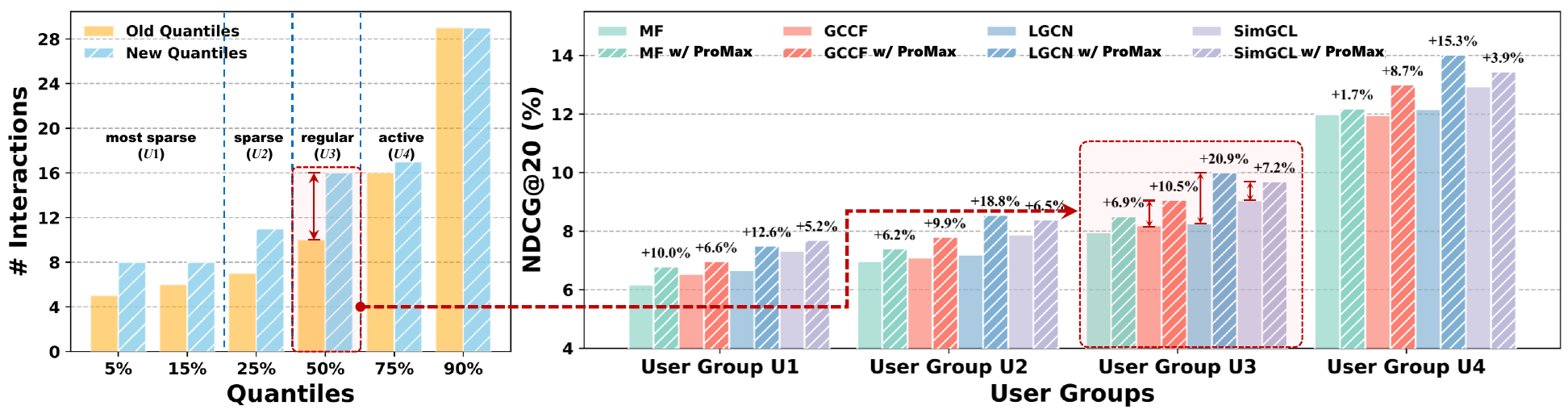}
\label{fig_yelp_sparsity}}

  \caption{Performance of \textsf{ProMax} and four base models (MF, GCCF, LightGCN, and SimGCL) on (a) Amazon-Book and (b) Yelp datasets under different levels of interaction sparsity.}
\label{fig_sparsity}
\end{figure}

Based on the direct performance comparison, we further investigate how \textsf{ProMax} enhances these traditional recommendation models. Specifically, the left side of Fig. \ref{fig_sparsity} presents the quantile statistics on the Amazon-Book and Yelp datasets before and after introducing sparsity-level shifts, while the right side reports the performance comparison between four base models and \textsf{ProMax} across the four sparsity groups. We observe that \textsf{ProMax} consistently outperforms all base models across all user groups, particularly among the less active groups. This indicates that the introduction of \textsf{ProMax} effectively alleviates the performance limitations caused by data sparsity. Notably, \textsf{ProMax} achieves the largest performance improvement in the regular user group (U3). We attribute this to the proposed sparsity-level shifting: as shown in the quantile statistics on the left side of Fig. \ref{fig_sparsity}, regular users gain the largest number of potential interactions after the shift, which substantially enriches their candidate item set and thus leads to significant performance gains.

\begin{figure}
\setlength{\abovecaptionskip}{0.0cm}
\setlength{\belowcaptionskip}{0.0cm} 
\centering
\subfigure[Amazon-Book]{\includegraphics[width=0.313\linewidth]{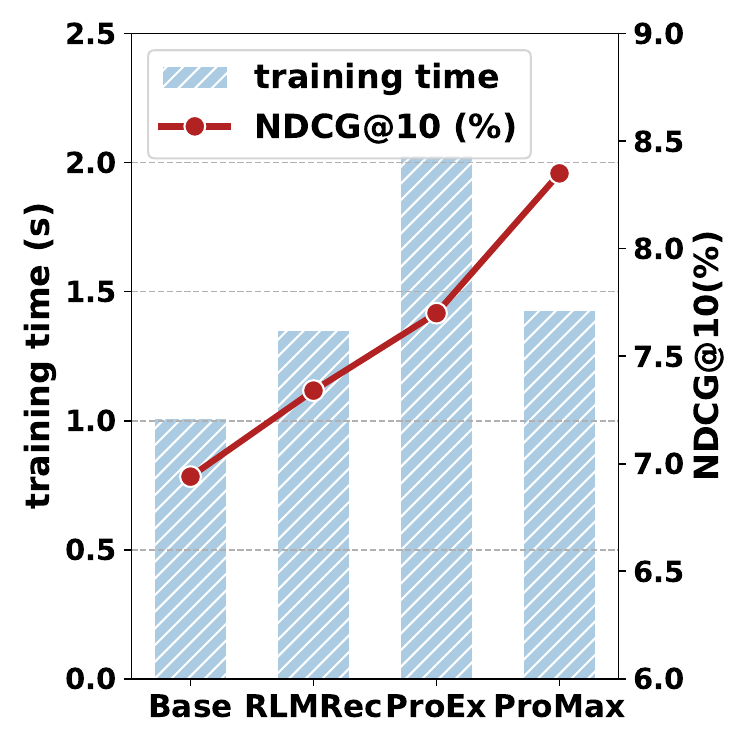}
\label{fig_amazon_time}}
\hfil
\subfigure[Yelp]{\includegraphics[width=0.313\linewidth]{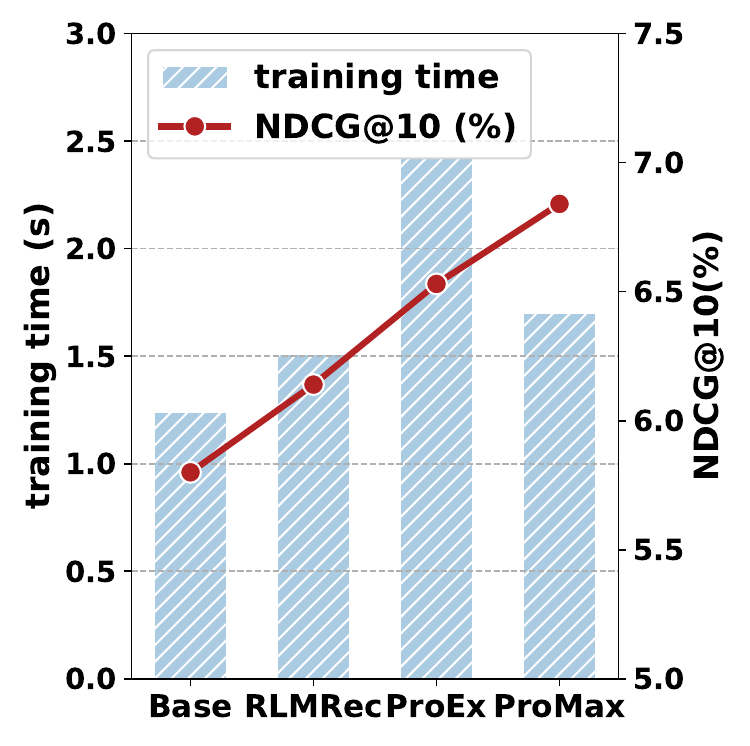}
\label{fig_yelp_time}}
\hfil
\subfigure[Steam]{\includegraphics[width=0.313\linewidth]{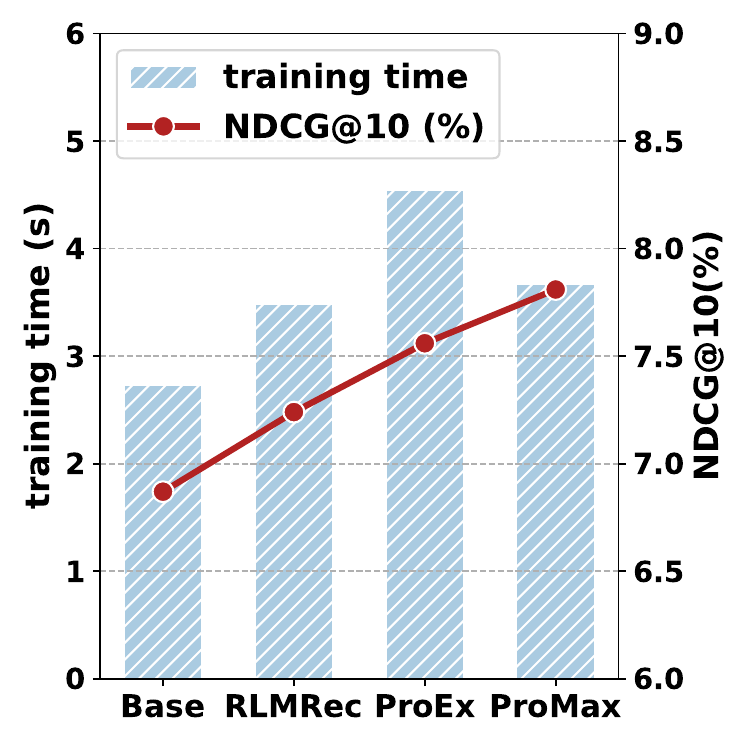}
\label{fig_steam_time}}
\caption{Comparison among \textsf{ProMax} and different methods \textit{w.r.t.} average runtime (left $y$-axis, lower is better) and recommendation performance (right $y$-axis, higher is better).}
\label{fig_time}
\end{figure}

\subsubsection{\textbf{Performance Comparison w.r.t. Efficiency.}}
\label{section_time}
Fig. \ref{fig_time} presents the comparison between computational efficiency and recommendation effectiveness across different methods on three datasets. As shown, the base model LightGCN consistently achieves the lowest runtime, but its recommendation performance is also the weakest. RLMRec introduces moderate computational overhead while yielding noticeable performance improvements, indicating that lightweight refinement already benefits recommendation quality. Compared with RLMRec, ProEx further improves NDCG@10 on all datasets, at the cost of increased runtime due to its enhanced exploration and multi-profiling strategy. Notably, \textsf{ProMax} consistently achieves the best recommendation performance across Amazon-Book, Yelp, and Steam datasets, while incurring only a moderate additional runtime overhead compared with base model. These results demonstrate that although \textsf{ProMax} introduces extra computation, its runtime remains within a practical range, and the performance gains are substantial and consistent across datasets.

\begin{figure}
\setlength{\abovecaptionskip}{0.0cm}
\setlength{\belowcaptionskip}{0.0cm} 
\centering
\subfigure[Amazon-Book]{\includegraphics[width=0.313\linewidth]{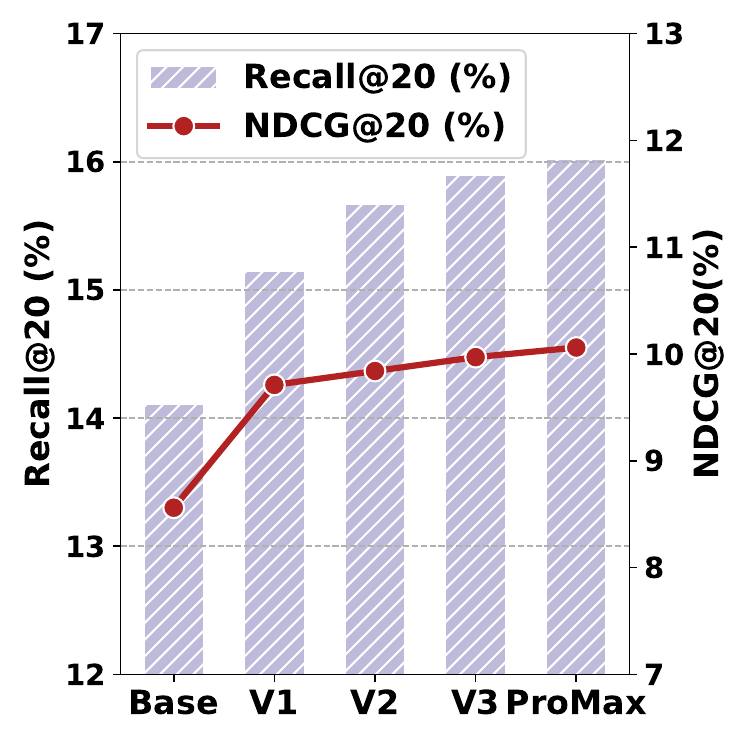}
\label{fig_amazon_ablation}}
\hfil
\subfigure[Yelp]{\includegraphics[width=0.313\linewidth]{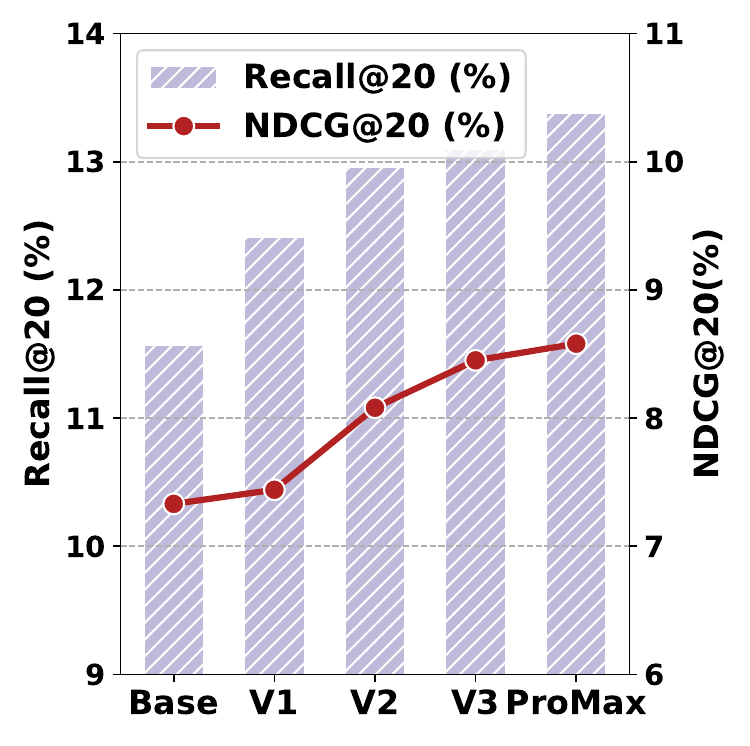}
\label{fig_yelp_ablation}}
\hfil
\subfigure[Steam]{\includegraphics[width=0.313\linewidth]{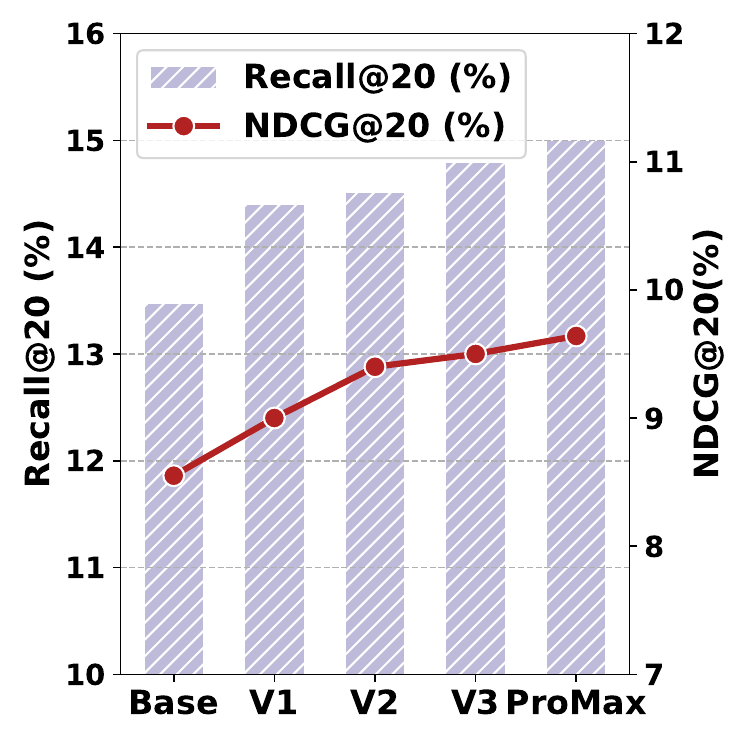}
\label{fig_steam__ablation}}
\caption{Ablation studies for \textsf{ProMax} on (a) Amazon-Book, (b) Yelp, and (c) Steam datasets \textit{w.r.t.} Recall@20 (left $y$-axis) and NDCG@20 (right $y$-axis).}
\label{fig_ablation}
\end{figure}

\subsection{In-depth Analysis of \textsf{ProMax}}
\subsubsection{\textbf{Ablation Studies}}
In this section, we develop multiple variants to validate the contribution of several core design choices underlying \textsf{ProMax}:
\begin{itemize}[leftmargin=*]
\item \textbf{V1}: remove the self-supervised distribution reshaping (Eq. \ref{cross_align});
\item \textbf{V2}: remove the supervised distribution reshaping (Eq. \ref{supervised_align});
\item \textbf{V3}: Replace the re-ranking item set $\mathcal R_u^{\text{LLM}}$ with the original candidate set $\mathcal R_u^{\text{item}}$ to obtain new interaction set $\mathcal M_u'$.
\end{itemize}

The experimental results for base model, all variants, and \textsf{ProMax} are shown in Fig. \ref{fig_ablation}. Firstly, all variants as well as \textsf{ProMax} achieve substantially higher performance than the base model, demonstrating the effectiveness of \textsf{ProMax}’s core components. Secondly, compared with the \textsf{ProMax}, all variants exhibit varying degrees of performance degradation, indicating the necessity of each module. This observation is consistent with our analysis: V1 relies solely on the original profiles, while V2 lacks the direct guidance provided by LLM-driven profiles. Only when both distribution reshaping processes are employed together can the full potential of the profiles be fully exploited. Finally,  to examine whether the LLM can contribute to improving the retrieval process, we feed the original candidate set $\mathcal R_u^{\text{item}}$ into \textsf{ProMax}. It should be noted that the sparsity-level shift strategy is likewise applied to $\mathcal R_u^{\text{item}}$ to ensure a fair comparison. It can be observed that \textsf{ProMax} equipped with $\mathcal{R}_u^{\text{LLM}}$ achieves slightly better performance than the original candidate set $\mathcal{R}_u^{\text{item}}$, which validates both the rationale and effectiveness of leveraging LLMs for fine-grained semantic matching.

\begin{figure}
\setlength{\abovecaptionskip}{0.0cm}
\setlength{\belowcaptionskip}{0.0cm} 
\centering
\subfigure[$\lambda_1$]{\includegraphics[width=0.48\linewidth]{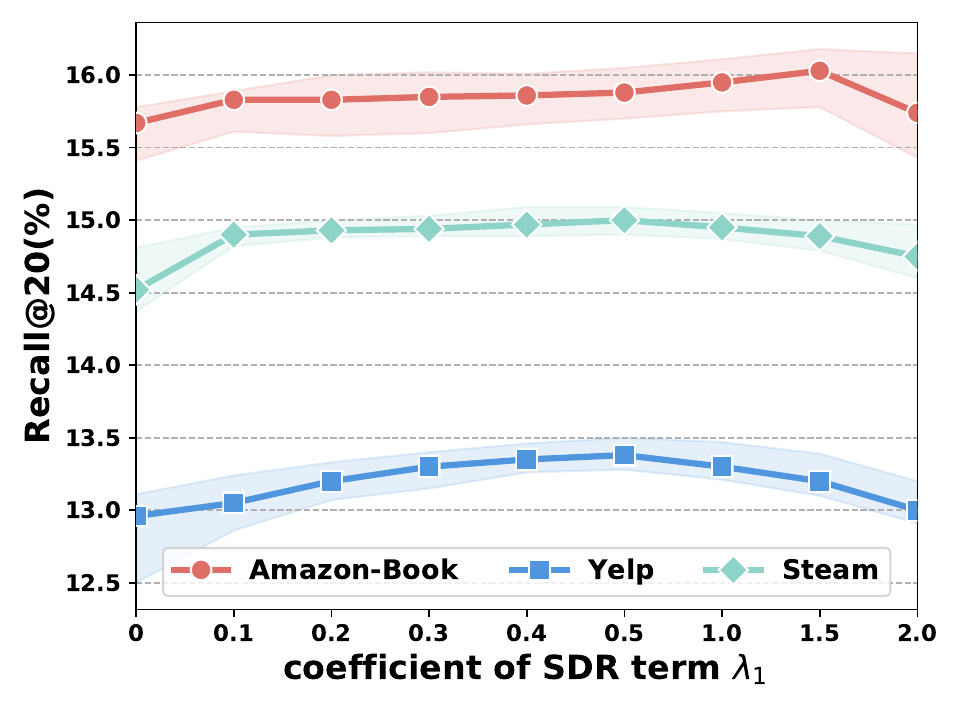}
\label{fig_lambda_1}}
\hfil
\subfigure[$\lambda_2$]{\includegraphics[width=0.48\linewidth]{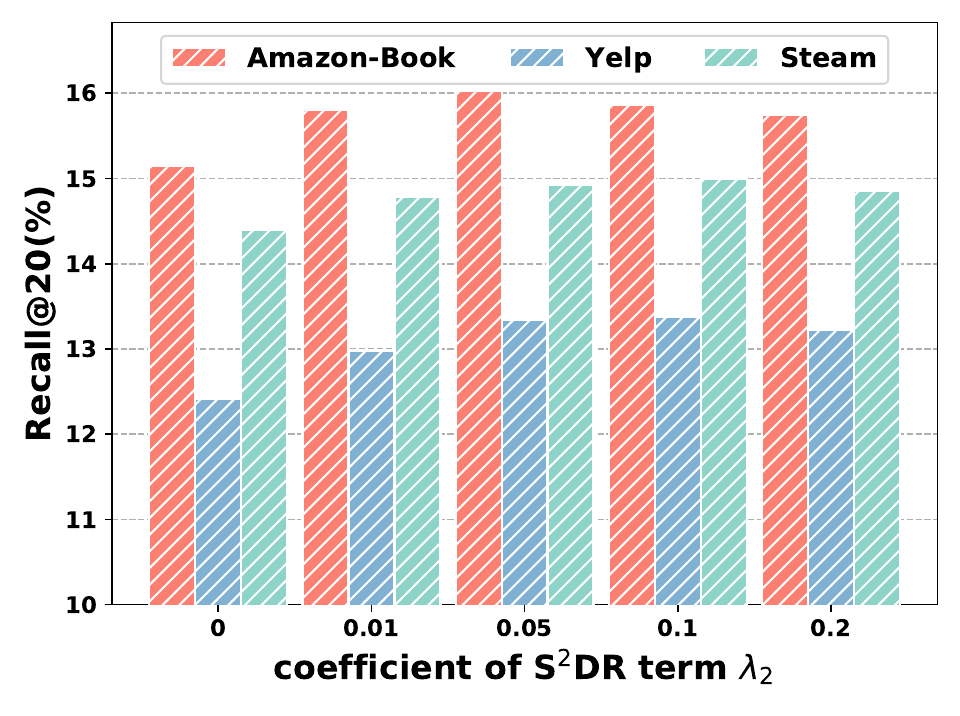}
\label{fig_lambda_2}}

\caption{Impact of hyperparameters (a) $\lambda_1$ and $\lambda_2$ for \textsf{ProMax} on Amazon-Book, Yelp, and Steam datasets \textit{w.r.t.} Recall@20.}
\label{fig_lambda}
\end{figure}

\subsubsection{\textbf{Impact of Hyperparameters}} Finally, we present the two new hyperparameters $\lambda_1$ and $\lambda_2$ introduced in the distribution reshaping process, as shown in Fig. \ref{fig_lambda}. We can observe that the optimal hyperparameters exhibit a similar trend across different datasets. Specifically, the optimal values of $\lambda_1$ on Amazon-Book, Yelp, and Steam datasets are 1.5, 0.5, and 0.5, respectively, while those of $\lambda_2$ are 0.05, 0.1, and 0.1, respectively. Furthermore, the Amazon-Book dataset exhibits a trend distinct from the other two datasets, which can be primarily attributed to the intrinsic differences in their data distributions. Fortunately, \textsf{ProMax} demonstrates strong robustness with respect to the two hyperparameters $\lambda_1$ and $\lambda_2$, such that its performance remains stable even when the optimal combination is not precisely identified. Similar observations are also made for NDCG and top-10 results, and are therefore omitted here for brevity.
\section{Related Work}

The powerful text processing capabilities of large language models \cite{zhao2023survey, xu2026vc} have spurred their growing integration into recommender systems, which has in turn attracted substantial research interest \cite{wu2024survey}. Existing methods generally fall into two categories. The first line of research generally incorporates the recommendation objective into the fine-tuning stage of large language models \cite{bao2023tallrec, wang2025multimodal}. For example, P5 \cite{geng2022recommendation} reformulates user–item interaction data into textual prompts to fine-tune the model. Subsequent studies such as TallRec \cite{bao2023tallrec} and LLaRA \cite{liao2024llara} integrate adapter modules or adopt LoRA \cite{hulora2022} to improve parameter efficiency during fine-tuning. More recent studies further incorporate direct preference optimization into the post-training stage \cite{gao2025sprec}. Despite recent progress, the heavy computational requirements and extended training time of fine-tuning approaches \cite{ren2024representation} often restrict their applicability to specific deployment scenarios \cite{liao2024llara, sheng2025language}. Specifically, while fine-tuning-based methods are mainly applied to sequential or session recommendation tasks \cite{zhao2025diversity}, they often fail to achieve competitive performance in general recommendation scenarios \cite{geng2022recommendation} requiring full ranking \cite{he2020lightgcn, yu2022graph}. This limitation suggests that relying solely on LLMs for recommendation or re-ranking remains insufficient \cite{ren2024representation, sheng2025language}.

The second line of research maintains the base recommendation model and integrates the LLM as an auxiliary module. For instance, multiple agents \cite{zhang2026smartagent} can be designed by exploiting the reasoning and memory capabilities of LLMs to support recommendation decision-making \cite{zhang2024agentcf}, whereas its text processing and generation abilities can be utilized to produce richer representations of users and items \cite{ren2024representation}. While the first category centers on formulating and evolving rules \cite{zhang2024generative}, the second prioritizes consistency within the feature space, focusing on transforming textual prompts into actionable latent representations (\textit{e.g}, profiles) \cite{xi2024towards, wang2025lettingo}. Recent works, including KAR \cite{xi2024towards}, RLMRec \cite{ren2024representation}, IRLLRec \cite{wang2025intent}, and ProEx \cite{zhang2025proex}, emphasize the critical role of aligning the representation spaces of recommendation models with those of LLMs \cite{radford2021learning, zhang2025towards}. Furthermore, some works focus on refining semantic representations through techniques like whitening \cite{zhang2024id} and compression \cite{hu2025alphafuse} to better align them with recommendation models. More recent approaches go a step further by directly substituting recommendation models’ ID embeddings with LLM-driven representations \cite{sheng2025language}. However, most existing studies have not thoroughly investigated the positive impact of semantic representations such as user or item profiles on recommendation. They merely treat such representations as auxiliary components and fail to fully exploit the potential of profiles.

\section{Conclusion}

In this paper, we revisited LLM-derived profiles for recommendation and proposed a simple yet effective model-agnostic framework \textsf{ProMax}. We first analyzed the collaborative signals revealed by profiles from a retrieval perspective. Subsequently, to maximize the potential of user and item profiles, they were used as indicators to guide the training of the recommendation model through a dual distribution reshaping process. In this process, the LLM-driven profiles serve only as indicators and do not participate in the training. Experiments on three datasets show that \textsf{ProMax} consistently improves the performance of diverse base recommendation models and outperforms existing LLM-enhanced methods.

\begin{acks}
This work is supported by the National Natural Science Foundation of China (No. 62272001), the Australian Research Council under the streams of Future Fellowship (No. FT210100624), the Discovery Early Career Researcher Award (No. DE230101033), the Discovery Project (No. DP240101108, DP240101814, and DP260100326), and the Linkage Projects (No. LP230200892 and LP240200546). 
\end{acks}

\bibliographystyle{ACM-Reference-Format}
\bibliography{sample-base}

@incollection{ricci2011introduction,
  title={Introduction to Recommender Systems Handbook},
  author={Ricci, Francesco and Rokach, Lior and Shapira, Bracha},
  booktitle={Recommender Systems Handbook},
  pages={1--35},
  year={2011},
  publisher={Springer}
}

@inproceedings{he2017neural,
  title={Neural collaborative filtering},
  author={He, Xiangnan and Liao, Lizi and Zhang, Hanwang and Nie, Liqiang and Hu, Xia and Chua, Tat-Seng},
  booktitle={Proceedings of the 26th International Conference on World Wide Web},
  pages={173--182},
  year={2017}
}

@inproceedings{he2020lightgcn,
  title={Lightgcn: Simplifying and powering graph convolution network for recommendation},
  author={He, Xiangnan and Deng, Kuan and Wang, Xiang and Li, Yan and Zhang, Yongdong and Wang, Meng},
  booktitle={Proceedings of the 43rd International ACM SIGIR Conference on Research and Development in Information Retrieval},
  pages={639--648},
  year={2020}
}

@inproceedings{rendle2009bpr,
  title={BPR: Bayesian Personalized Ranking from Implicit Feedback},
  author={Rendle, Steffen and Freudenthaler, Christoph and Gantner, Zeno and Schmidt-Thieme, Lars},
  booktitle={Proceedings of the Twenty-Fifth Conference on Uncertainty in Artificial Intelligence (UAI)},
  pages={452--461},
  year={2009}
}

@inproceedings{rendle2020neural,
  title={Neural collaborative filtering vs. matrix factorization revisited},
  author={Rendle, Steffen and Krichene, Walid and Zhang, Li and Anderson, John},
  booktitle={Proceedings of the 14th ACM Conference on Recommender Systems},
  pages={240--248},
  year={2020}
}

@inproceedings{yuan2023go,
  title={Where to go next for recommender systems? id-vs. modality-based recommender models revisited},
  author={Yuan, Zheng and Yuan, Fajie and Song, Yu and Li, Youhua and Fu, Junchen and Yang, Fei and Pan, Yunzhu and Ni, Yongxin},
  booktitle={Proceedings of the 46th International ACM SIGIR Conference on Research and Development in Information Retrieval},
  pages={2639--2649},
  year={2023}
}

@inproceedings{yu2022graph,
  title={Are graph augmentations necessary? simple graph contrastive learning for recommendation},
  author={Yu, Junliang and Yin, Hongzhi and Xia, Xin and Chen, Tong and Cui, Lizhen and Nguyen, Quoc Viet Hung},
  booktitle={Proceedings of the 45th International ACM SIGIR Conference on Research and Development in Information Retrieval},
  pages={1294--1303},
  year={2022}
}

@article{chang2024survey,
  title={A survey on evaluation of large language models},
  author={Chang, Yupeng and Wang, Xu and Wang, Jindong and Wu, Yuan and Yang, Linyi and Zhu, Kaijie and Chen, Hao and Yi, Xiaoyuan and Wang, Cunxiang and Wang, Yidong and others},
  journal={ACM Transactions on Intelligent Systems and Technology},
  volume={15},
  number={3},
  pages={1--45},
  year={2024},
  publisher={ACM New York, NY}
}

@article{zhao2023survey,
  title={A survey of large language models},
  author={Zhao, Wayne Xin and Zhou, Kun and Li, Junyi and Tang, Tianyi and Wang, Xiaolei and Hou, Yupeng and Min, Yingqian and Zhang, Beichen and Zhang, Junjie and Dong, Zican and others},
  journal={arXiv preprint arXiv:2303.18223},
  year={2023}
}

@inproceedings{liao2024llara,
  title={Llara: Large language-recommendation assistant},
  author={Liao, Jiayi and Li, Sihang and Yang, Zhengyi and Wu, Jiancan and Yuan, Yancheng and Wang, Xiang and He, Xiangnan},
  booktitle={Proceedings of the 47th International ACM SIGIR Conference on Research and Development in Information Retrieval},
  pages={1785--1795},
  year={2024}
}

@inproceedings{gao2025sprec,
  title={Sprec: Self-play to debias llm-based recommendation},
  author={Gao, Chongming and Chen, Ruijun and Yuan, Shuai and Huang, Kexin and Yu, Yuanqing and He, Xiangnan},
  booktitle={Proceedings of the ACM on Web Conference 2025},
  pages={5075--5084},
  year={2025}
}

@inproceedings{wei2024llmrec,
  title={Llmrec: Large language models with graph augmentation for recommendation},
  author={Wei, Wei and Ren, Xubin and Tang, Jiabin and Wang, Qinyong and Su, Lixin and Cheng, Suqi and Wang, Junfeng and Yin, Dawei and Huang, Chao},
  booktitle={Proceedings of the 17th ACM International Conference on Web Search and Data Mining},
  pages={806--815},
  year={2024}
}

@inproceedings{xi2024towards,
  title={Towards open-world recommendation with knowledge augmentation from large language models},
  author={Xi, Yunjia and Liu, Weiwen and Lin, Jianghao and Cai, Xiaoling and Zhu, Hong and Zhu, Jieming and Chen, Bo and Tang, Ruiming and Zhang, Weinan and Yu, Yong},
  booktitle={Proceedings of the 18th ACM Conference on Recommender Systems},
  pages={12--22},
  year={2024}
}

@inproceedings{wang2025lettingo,
  title={LettinGo: Explore User Profile Generation for Recommendation System},
  author={Wang, Lu and Zhang, Di and Yang, Fangkai and Zhao, Pu and Liu, Jianfeng and Zhan, Yuefeng and Sun, Hao and Lin, Qingwei and Deng, Weiwei and Zhang, Dongmei and others},
  booktitle={Proceedings of the 31st ACM SIGKDD Conference on Knowledge Discovery and Data Mining V. 2},
  pages={2985--2995},
  year={2025}
}

@inproceedings{ren2024representation,
  title={Representation learning with large language models for recommendation},
  author={Ren, Xubin and Wei, Wei and Xia, Lianghao and Su, Lixin and Cheng, Suqi and Wang, Junfeng and Yin, Dawei and Huang, Chao},
  booktitle={Proceedings of the ACM on Web Conference 2024},
  pages={3464--3475},
  year={2024}
}

@inproceedings{wang2025intent,
  title={Intent representation learning with large language model for recommendation},
  author={Wang, Yu and Sang, Lei and Zhang, Yi and Zhang, Yiwen},
  booktitle={Proceedings of the 48th International ACM SIGIR Conference on Research and Development in Information Retrieval},
  pages={1870--1879},
  year={2025}
}

@inproceedings{hu2025alphafuse,
  title={Alphafuse: Learn id embeddings for sequential recommendation in null space of language embeddings},
  author={Hu, Guoqing and Zhang, An and Liu, Shuo and Cai, Zhibo and Yang, Xun and Wang, Xiang},
  booktitle={Proceedings of the 48th International ACM SIGIR Conference on Research and Development in Information Retrieval},
  pages={1614--1623},
  year={2025}
}

@inproceedings{zhang2024id,
  title={Are id embeddings necessary? whitening pre-trained text embeddings for effective sequential recommendation},
  author={Zhang, Lingzi and Zhou, Xin and Zeng, Zhiwei and Shen, Zhiqi},
  booktitle={2024 IEEE 40th International Conference on Data Engineering (ICDE)},
  pages={530--543},
  year={2024},
  organization={IEEE}
}

@inproceedings{sheng2025language,
  title={Language Representations Can be What Recommenders Need: Findings and Potentials},
  author={Sheng, Leheng and Zhang, An and Zhang, Yi and Chen, Yuxin and Wang, Xiang and Chua, Tat-Seng},
  booktitle={The Thirteenth International Conference on Learning Representations},
  year={2025}
}

@inproceedings{wang2025unleashing,
  title={Unleashing the Power of Large Language Model for Denoising Recommendation},
  author={Wang, Shuyao and Zheng, Zhi and Sui, Yongduo and Xiong, Hui},
  booktitle={Proceedings of the ACM on Web Conference 2025},
  pages={252--263},
  year={2025}
}

@inproceedings{zhang2022effect,
  title={On the effect of isotropy on VAE representations of text},
  author={Zhang, Lan and Buntine, Wray and Shareghi, Ehsan},
  booktitle={Annual Meeting of the Association of Computational Linguistics 2022},
  pages={694--701},
  year={2022},
  organization={Association for Computational Linguistics (ACL)}
}

@article{sobal2024mathbb,
  title={X-Sample Contrastive Loss: Improving Contrastive Learning with Sample Similarity Graphs},
  author={Sobal, Vlad and Ibrahim, Mark and Balestriero, Randall and Cabannes, Vivien and Bouchacourt, Diane and Astolfi, Pietro and Cho, Kyunghyun and LeCun, Yann},
  journal={arXiv preprint arXiv:2407.18134},
  year={2024}
}

@inproceedings{chen2020revisiting,
  title={Revisiting graph based collaborative filtering: A linear residual graph convolutional network approach},
  author={Chen, Lei and Wu, Le and Hong, Richang and Zhang, Kun and Wang, Meng},
  booktitle={Proceedings of the AAAI Conference on Artificial Intelligence},
  volume={34},
  number={01},
  pages={27--34},
  year={2020}
}

@inproceedings{wang2024collaborative,
  title={Collaborative alignment for recommendation},
  author={Wang, Chen and Yang, Liangwei and Liu, Zhiwei and Liu, Xiaolong and Yang, Mingdai and Liang, Yueqing and Yu, Philip S},
  booktitle={Proceedings of the 33rd ACM International Conference on Information and Knowledge Management},
  pages={2315--2325},
  year={2024}
}

@article{liu2024llm,
  title={Llm-esr: Large language models enhancement for long-tailed sequential recommendation},
  author={Liu, Qidong and Wu, Xian and Wang, Yejing and Zhang, Zijian and Tian, Feng and Zheng, Yefeng and Zhao, Xiangyu},
  journal={Advances in Neural Information Processing Systems},
  volume={37},
  pages={26701--26727},
  year={2024}
}

@article{wu2024survey,
  title={A survey on large language models for recommendation},
  author={Wu, Likang and Zheng, Zhi and Qiu, Zhaopeng and Wang, Hao and Gu, Hongchao and Shen, Tingjia and Qin, Chuan and Zhu, Chen and Zhu, Hengshu and Liu, Qi and others},
  journal={World Wide Web},
  volume={27},
  number={5},
  pages={60},
  year={2024},
  publisher={Springer}
}

@inproceedings{fang2024scaling,
  title={Scaling laws for dense retrieval},
  author={Fang, Yan and Zhan, Jingtao and Ai, Qingyao and Mao, Jiaxin and Su, Weihang and Chen, Jia and Liu, Yiqun},
  booktitle={Proceedings of the 47th International ACM SIGIR Conference on Research and Development in Information Retrieval},
  pages={1339--1349},
  year={2024}
}

@inproceedings{kabbur2013fism,
  title={Fism: factored item similarity models for top-n recommender systems},
  author={Kabbur, Santosh and Ning, Xia and Karypis, George},
  booktitle={Proceedings of the 19th ACM SIGKDD international conference on Knowledge discovery and data mining},
  pages={659--667},
  year={2013}
}

@article{zhang2023revisiting,
  title={Revisiting graph-based recommender systems from the perspective of variational auto-encoder},
  author={Zhang, Yi and Zhang, Yiwen and Yan, Dengcheng and Deng, Shuiguang and Yang, Yun},
  journal={ACM Transactions on Information Systems},
  volume={41},
  number={3},
  pages={1--28},
  year={2023},
  publisher={ACM New York, NY}
}

@inproceedings{kang2018self,
  title={Self-attentive sequential recommendation},
  author={Kang, Wang-Cheng and McAuley, Julian},
  booktitle={2018 IEEE international conference on data mining (ICDM)},
  pages={197--206},
  year={2018},
  organization={IEEE}
}

@inproceedings{zhang2024agentcf,
  title={Agentcf: Collaborative learning with autonomous language agents for recommender systems},
  author={Zhang, Junjie and Hou, Yupeng and Xie, Ruobing and Sun, Wenqi and McAuley, Julian and Zhao, Wayne Xin and Lin, Leyu and Wen, Ji-Rong},
  booktitle={Proceedings of the ACM Web Conference 2024},
  pages={3679--3689},
  year={2024}
}

@inproceedings{hulora2022,
  title={LoRA: Low-Rank Adaptation of Large Language Models},
  author={Hu, Edward J and Wallis, Phillip and Allen-Zhu, Zeyuan and Li, Yuanzhi and Wang, Shean and Wang, Lu and Chen, Weizhu and others},
  booktitle={International Conference on Learning Representations},
  year={2022}
}

@inproceedings{bao2023tallrec,
  title={Tallrec: An effective and efficient tuning framework to align large language model with recommendation},
  author={Bao, Keqin and Zhang, Jizhi and Zhang, Yang and Wang, Wenjie and Feng, Fuli and He, Xiangnan},
  booktitle={Proceedings of the 17th ACM Conference on Recommender Systems},
  pages={1007--1014},
  year={2023}
}

@article{kingma2014adam,
  title={Adam: A Method for Stochastic Optimization},
  author={Kingma, Diederik P and Ba, Jimmy},
  journal={arXiv preprint arXiv:1412.6980},
  year={2014}
}

@inproceedings{glorot2010understanding,
  title={Understanding the Difficulty of Training Deep Feedforward Neural Networks},
  author={Glorot, Xavier and Bengio, Yoshua},
  booktitle={Proceedings of the Thirteenth International Conference on Artificial Intelligence and Statistics (ICAIS)},
  pages={249--256},
  year={2010}
}

@inproceedings{geng2022recommendation,
  title={Recommendation as language processing (rlp): A unified pretrain, personalized prompt \& predict paradigm (p5)},
  author={Geng, Shijie and Liu, Shuchang and Fu, Zuohui and Ge, Yingqiang and Zhang, Yongfeng},
  booktitle={Proceedings of the 16th ACM Conference on Recommender Systems},
  pages={299--315},
  year={2022}
}

@inproceedings{zhang2024generative,
  title={On generative agents in recommendation},
  author={Zhang, An and Chen, Yuxin and Sheng, Leheng and Wang, Xiang and Chua, Tat-Seng},
  booktitle={Proceedings of the 47th international ACM SIGIR conference on research and development in Information Retrieval},
  pages={1807--1817},
  year={2024}
}

@inproceedings{radford2021learning,
  title={Learning transferable visual models from natural language supervision},
  author={Radford, Alec and Kim, Jong Wook and Hallacy, Chris and Ramesh, Aditya and Goh, Gabriel and Agarwal, Sandhini and Sastry, Girish and Askell, Amanda and Mishkin, Pamela and Clark, Jack and others},
  booktitle={International Conference on Machine Learning},
  pages={8748--8763},
  year={2021},
  organization={PMLR}
}

@inproceedings{zhang2025towards,
  title={Towards distribution matching between collaborative and language spaces for generative recommendation},
  author={Zhang, Yi and Zhang, Yiwen and Wang, Yu and Chen, Tong and Yin, Hongzhi},
  booktitle={Proceedings of the 48th International ACM SIGIR Conference on Research and Development in Information Retrieval},
  pages={2006--2016},
  year={2025}
}

@article{yu2023self,
  title={Self-supervised learning for recommender systems: A survey},
  author={Yu, Junliang and Yin, Hongzhi and Xia, Xin and Chen, Tong and Li, Jundong and Huang, Zi},
  journal={IEEE Transactions on Knowledge and Data Engineering},
  volume={36},
  number={1},
  pages={335--355},
  year={2023},
  publisher={IEEE}
}

@inproceedings{wang2019neural,
  title={Neural graph collaborative filtering},
  author={Wang, Xiang and He, Xiangnan and Wang, Meng and Feng, Fuli and Chua, Tat-Seng},
  booktitle={Proceedings of the 42nd international ACM SIGIR Conference on Research and development in Information Retrieval},
  pages={165--174},
  year={2019}
}

@article{chen2023bias,
  title={Bias and debias in recommender system: A survey and future directions},
  author={Chen, Jiawei and Dong, Hande and Wang, Xiang and Feng, Fuli and Wang, Meng and He, Xiangnan},
  journal={ACM Transactions on Information Systems},
  volume={41},
  number={3},
  pages={1--39},
  year={2023},
  publisher={ACM New York, NY}
}

@article{ji2023survey,
  title={Survey of hallucination in natural language generation},
  author={Ji, Ziwei and Lee, Nayeon and Frieske, Rita and Yu, Tiezheng and Su, Dan and Xu, Yan and Ishii, Etsuko and Bang, Ye Jin and Madotto, Andrea and Fung, Pascale},
  journal={ACM Computing Surveys},
  volume={55},
  number={12},
  pages={1--38},
  year={2023},
  publisher={ACM New York, NY}
}

@inproceedings{chen2023adap,
  title={Adap-$\tau$: Adaptively modulating embedding magnitude for recommendation},
  author={Chen, Jiawei and Wu, Junkang and Wu, Jiancan and Cao, Xuezhi and Zhou, Sheng and He, Xiangnan},
  booktitle={Proceedings of the ACM Web Conference 2023},
  pages={1085--1096},
  year={2023}
}

@article{shannon2001mathematical,
  title={A mathematical theory of communication},
  author={Shannon, Claude E},
  journal={ACM SIGMOBILE mobile computing and communications review},
  volume={5},
  number={1},
  pages={3--55},
  year={2001},
  publisher={ACM New York, NY, USA}
}

@inproceedings{zhang2025proex,
  title={ProEx: A Unified Framework Leveraging Large Language Model with Profile Extrapolation for Recommendation},
  author={Zhang, Yi and Zhang, Yiwen and Wang, Yu and Chen, Tong and Yin, Hongzhi},
  booktitle={Proceedings of the 32nd ACM
SIGKDD Conference on Knowledge Discovery and Data Mining V.1},
  year={2026}
}

@inproceedings{xu2026multi,
  title={Multi-Value Alignment for LLMs via Value Decorrelation and Extrapolation},
  author={Xu, Hefei and Wu, Le and Cheng, Chen and Liu, Hao},
  booktitle={Proceedings of the AAAI Conference on Artificial Intelligence},
  volume={40},
  number={40},
  pages={34133--34141},
  year={2026}
}

@inproceedings{xu2026vc,
  title={VC-Soup: Value-Consistency Guided Multi-Value Alignment for Large Language Models},
  author={Xu, Hefei and Wu, Le and Wang, Yu and Hou, Min and Wu, Han and Zhang, Zhen and Wang, Meng},
  booktitle={Proceedings of the ACM Web Conference 2026},
  pages={9699--9710},
  year={2026}
}

@inproceedings{zhang_lightccf_2025,
title = {Unveiling Contrastive Learning‘ Capability of Neighborhood Aggregation for Collaborative Filtering},
author={Zhang, Yu and Zhang, Yiwen and Zhang, Yi and Sang, Lei and Yang, Yun},
booktitle = {Proceedings of the 48th International ACM SIGIR Conference on Research and Development in Information Retrieval},
pages = {1985–1994},
year={2025},
}

@article{wang2026mllmrec,
  title={MLLMRec-R1: Incentivizing Reasoning Capability in Large Language Models for Multimodal Sequential Recommendation},
  author={Wang, Yu and Yang, Yonghui and Wu, Le and Wu, Jiancan and Xu, Hefei and Lin, Hui},
  journal={arXiv preprint arXiv:2603.06243},
  year={2026}
}

@article{wang2025multimodal,
  title={Multimodal Large Language Models with Adaptive Preference Optimization for Sequential Recommendation},
  author={Wang, Yu and Yang, Yonghui and Wu, Le and Zhang, Yi and Hong, Richang},
  journal={arXiv preprint arXiv:2511.18740},
  year={2025}
}

@article{yin2025device,
  title={On-device recommender systems: A comprehensive survey},
  author={Yin, Hongzhi and Qu, Liang and Chen, Tong and Yuan, Wei and Zheng, Ruiqi and Long, Jing and Xia, Xin and Shi, Yuhui and Zhang, Chengqi},
  journal={Data Science and Engineering},
  pages={1--30},
  year={2025},
  publisher={Springer}
}

@inproceedings{zhao2025diversity,
  title={Diversity-aware Dual-promotion Poisoning Attack on Sequential Recommendation},
  author={Zhao, Yuchuan and Chen, Tong and Yu, Junliang and Zheng, Kai and Cui, Lizhen and Yin, Hongzhi},
  booktitle={Proceedings of the 48th International ACM SIGIR Conference on Research and Development in Information Retrieval},
  pages={1634--1644},
  year={2025}
}

@inproceedings{yin2015joint,
  title={Joint modeling of users' interests and mobility patterns for point-of-interest recommendation},
  author={Yin, Hongzhi and Cui, Bin and Huang, Zi and Wang, Weiqing and Wu, Xian and Zhou, Xiaofang},
  booktitle={Proceedings of the 23rd ACM international conference on Multimedia},
  pages={819--822},
  year={2015}
}

@inproceedings{zhang2026smartagent,
  title={Smartagent: Chain-of-user-thought for embodied personalized agent in cyber world},
  author={Zhang, Jiaqi and Gao, Chen and Zhang, Liyuan and Nguyen, Quoc Viet Hung and Yin, Hongzhi},
  booktitle={Proceedings of the AAAI Conference on Artificial Intelligence},
  volume={40},
  number={21},
  pages={17993--18001},
  year={2026}
}


\end{document}